# Tests of Lorentz invariance at the Sudbury Neutrino Observatory

B. Aharmim,[8] S. N. Ahmed,[16] A. E. Anthony,[18,b] N. Barros,[10,c] E. W. Beier,[15] A. Bellerive,[5] B. Beltran,[1] M. Bergevin,[9,7,d] S. D. Biller,[14] E. Blucher,[6] R. Bonventre,[2,9] K. Boudjemline,[5,16] M. G. Boulay,[16,e] B. Cai,[16] E. J. Callaghan,[2,9] J. Caravaca,[2,9] Y. D. Chan,[9] D. Chauhan,[8,f] M. Chen,[16] B. T. Cleveland,[14] G. A. Cox,[20,g] X. Dai,[16,14,5] H. Deng,[15,h] F. B. Descamps,[2,9] J. A. Detwiler,[9,i] P. J. Doe,[20] G. Doucas,[14] P.-L. Drouin,[5] M. Dunford,[15,j] S. R. Elliott,[11,20] H. C. Evans,[16,a] G. T. Ewan,[16] J. Farine,[8,5] H. Fergani,[14] F. Fleurot,[8] R. J. Ford,[17,16] J. A. Formaggio,[13,20] N. Gagnon,[20,11,9,14] K. Gilje,[1] J. TM. Goon,[12] K. Graham,[5,16] E. Guillian,[16] S. Habib,[1] R. L. Hahn,[4] A. L. Hallin,[1] E. D. Hallman,[8] P. J. Harvey,[16] R. Hazama,[20,k] W. J. Heintzelman,[15] J. Heise,[3,11,16,l] R. L. Helmer,[19] A. Hime,[11] C. Howard,[1] M. Huang,[18,8] P. Jagam,[7] B. Jamieson,[3,m] N. A. Jelley,[14] M. Jerkins,[18] C. Kéfélian,[2,9] K. J. Keeter,[17,n] J. R. Klein,[18,15] L. L. Kormos,[16,o] M. Kos,[16,p] A. Krüger,[8] C. Kraus,[16,8] C. B. Krauss,[1] T. Kutter,[12] C. C. M. Kyba,[15,q] K. Labe,[6,r] B. J. Land,[2,9] R. Lange,[4] A. LaTorre,[6] J. Law,[7] I. T. Lawson,[17,7] K. T. Lesko,[9] J. R. Leslie,[16] I. Levine,[5,s] J. C. Loach,[14,9] R. MacLellan,[16,t] S. Majerus,[14] H. B. Mak,[16] J. Maneira,[10] R. D. Martin,[16,9] A. Mastbaum,[6,15] N. McCauley,[15,14,u] A. B. McDonald,[16] S. R. McGee,[20] M. L. Miller,[13,i] B. Monreal,[13,v] J. Monroe,[13,w] B. G. Nickel,[7] A. J. Noble,[16,5] H. M. O'Keeffe,[14,o] N. S. Oblath,[20,13,x] C. E. Okada,[9,y] R. W. Ollerhead,[7] G. D. Orebi Gann,[2,15,9] S. M. Oser,[3,19] R. A. Ott,[13,z] S. J. M. Peeters,[14,aa] A. W. P. Poon,[9] G. Prior,[9,bb] S. D. Reitzner,[7,cc] K. Rielage,[11,20] B. C. Robertson,[16] R. G. H. Robertson,[20] M. H. Schwendener,[8] J. A. Secrest,[15,dd] S. R. Seibert,[18,11,15,ee] O. Simard,[5,ff] D. Sinclair,[5,19] P. Skensved,[16] T. J. Sonley,[13,f] L. C. Stonehill,[11,20] G. Tešić,[5,gg] N. Tolich,[20] T. Tsui,[3,hh] R. Van Berg,[15] B. A. VanDevender,[20,x] C. J. Virtue,[8] B. L. Wall,[20] D. Waller,[5] H. Wan Chan Tseung,[14,20] D. L. Wark,[14,ii] J. Wendland,[3] N. West,[14] J. F. Wilkerson,[20,jj] T. Winchester,[20] J. R. Wilson,[14,kk] A. Wright,[16] M. Yeh,[4] F. Zhang,[5,ll] and K. Zuber[14,mm]

(SNO Collaboration)

[1]*Department of Physics, University of Alberta, Edmonton, Alberta T6G 2R3, Canada*
[2]*Physics Department, University of California at Berkeley, Berkeley, California 94720-7300, USA*
[3]*Department of Physics and Astronomy, University of British Columbia, Vancouver, Brisith Columbia V6T 1Z1, Canada*
[4]*Chemistry Department, Brookhaven National Laboratory, Upton, New York 11973-5000, USA*
[5]*Ottawa-Carleton Institute for Physics, Department of Physics, Carleton University, Ottawa, Ontario K1S 5B6, Canada*
[6]*Department of Physics, University of Chicago, Chicago, Illinois 60637, USA*
[7]*Physics Department, University of Guelph, Guelph, Ontario N1G 2W1, Canada*
[8]*Department of Physics and Astronomy, Laurentian University, Sudbury, Ontario P3E 2C6, Canada*
[9]*Institute for Nuclear and Particle Astrophysics and Nuclear Science Division, Lawrence Berkeley National Laboratory, Berkeley, California 94720-8153, USA*
[10]*Laboratório de Instrumentação e Física Experimental de Partículas, Av. Elias Garcia 14, 1°, 1000-149 Lisboa, Portugal*
[11]*Los Alamos National Laboratory, Los Alamos, New Mexico 87545, USA*
[12]*Department of Physics and Astronomy, Louisiana State University, Baton Rouge, Louisiana 70803, USA*
[13]*Laboratory for Nuclear Science, Massachusetts Institute of Technology, Cambridge, Massachusetts 02139, USA*
[14]*Department of Physics, University of Oxford, Denys Wilkinson Building, Keble Road, Oxford OX1 3RH, United Kingdom*
[15]*Department of Physics and Astronomy, University of Pennsylvania, Philadelphia, Pennsylvania 19104-6396, USA*
[16]*Department of Physics, Queen's University, Kingston, Ontario K7L 3N6, Canada*
[17]*SNOLAB, Lively, Ontario P3Y 1N2, Canada*
[18]*Department of Physics, University of Texas at Austin, Austin, Texas 78712-0264, USA*

[19]TRIUMF, 4004 Wesbrook Mall, Vancouver, British Columbia V6T 2A3, Canada
[20]Center for Experimental Nuclear Physics and Astrophysics, and Department of Physics, University of Washington, Seattle, Washington 98195, USA

Experimental tests of Lorentz symmetry in systems of all types are critical for ensuring that the basic assumptions of physics are well founded. Data from all phases of the Sudbury Neutrino Observatory, a kiloton-scale heavy water Cherenkov detector, are analyzed for possible violations of Lorentz symmetry in the neutrino sector. Such violations would appear as one of eight possible signal types in the detector: six seasonal variations in the solar electron neutrino survival probability differing in energy and time dependence and two shape changes to the oscillated solar neutrino energy spectrum. No evidence for such signals is observed, and limits on the size of such effects are established in the framework of the standard model extension, including 38 limits on previously unconstrained operators and improved limits on 16 additional operators. This makes limits on all minimal, Dirac-type Lorentz violating operators in the neutrino sector available for the first time.

[a]Deceased.
[b]Preset address: Global Development Lab, U.S. Agency for International Development, Washington, DC, USA.
[c]Present address: Department of Physics and Astronomy, University of Pennsylvania, Philadelphia, Pennsylvania, USA.
[d]Present address: Lawrence Livermore National Laboratory, Livermore, California, USA.
[e]Present address: Department of Physics, Carleton University, Ottawa, Ontario, Canada.
[f]Present address: SNOLAB, Lively, Ontario, Canada.
[g]Present address: Institut für Experimentelle Kernphysik, Karlsruher Institut für Technologie, Karlsruhe, Germany.
[h]Present address: Rock Creek Group, Washington, DC, USA.
[i]Present address: Center for Experimental Nuclear Physics and Astrophysics, and Department of Physics, University of Washington, Seattle, Washington, USA.
[j]Present address: Ruprecht-Karls-Universität Heidelberg, Heidelberg, Germany.
[k]Present address: Research Center for Nuclear Physics, Osaka, Japan.
[l]Present address: Sanford Underground Research Laboratory, Lead, South Dakota, USA.
[m]Present address: Department of Physics, University of Winnipeg, Winnipeg, Manitoba, Canada.
[n]Present address: Black Hills State University, Spearfish, South Dakota, USA.
[o]Present address: Physics Department, Lancaster University, Lancaster, United Kingdom.
[p]Present address: Pelmorex Corp., Oakville, Ontario, Canada.
[q]Present address: GFZ German Research Centre for Geosciences, Potsdam, Germany.
[r]Present address: Department of Physics, Cornell University, Ithaca, New York, USA.
[s]Present address: Department of Physics and Astronomy, Indiana University, South Bend, Indiana, USA.
[t]Present address: University of South Dakota, Vermillion, South Dakota, USA.
[u]Present address: Department of Physics, University of Liverpool, Liverpool, United Kingdom.
[v]Present address: Department of Physics, Case Western Reserve University, Cleveland, Ohio, USA.
[w]Present address: Dept. of Physics, Royal Holloway University of London, Egham, Surrey, United Kingdom.
[x]Present address: Pacific Northwest National Laboratory, Richland, Washington, USA.
[y]Present address: Nevada National Security Site, Las Vegas, Nevada, USA.
[z]Present address: Department of Physics, University of California, Davis, California, USA.
[aa]Present address: Department of Physics and Astronomy, University of Sussex, Brighton, United Kingdom.
[bb]Present address: Laboratório de Instrumentação e Física Experimental de Partículas, Lisboa, Portugal.
[cc]Present address: Fermilab, Batavia, Illinois, USA.
[dd]Present address: Dept. of Physics, Georgia Southern University, Statesboro, Georgia, USA.
[ee]Present address: Continuum Analytics, Austin, Texas, USA.
[ff]Present address: CEA-Saclay, DSM/IRFU/SPP, Gif-sur-Yvette, France.
[gg]Present address: Physics Department, McGill University, Montreal, Quebec, Canada.
[hh]Present address: Kwantlen Polytechnic University, Surrey, British Columbia, Canada.
[ii]Additional address: Rutherford Appleton Laboratory, Chilton, Didcot, United Kingdom.
[jj]Present address: Department of Physics, University of North Carolina, Chapel Hill, North Carolina, USA.
[kk]Present address: Dept. of Physics, Queen Mary University, London, United Kingdom.
[ll]Present address: Laufer Center, Stony Brook University, Stony Brook, New York, USA.
[mm]Present address: Institut für Kern- und Teilchenphysik, Technische Universität Dresden, Dresden, Germany.

## I. INTRODUCTION

Solar neutrinos are produced in the electron flavor. At the relevant energies, the electron flavor fraction of the active solar neutrino flux after it has propagated to the Earth is roughly 1/3. The Sudbury Neutrino Observatory (SNO) [1] was able to make a precise measurement of this flavor fraction through its distinct flavor-tagging [2] and flavor-neutral [3] detection channels.

Lorentz symmetry is one of the underlying assumptions on which the standard model of particle physics is built. However, the degree to which this symmetry is respected is an experimental question, and searches for its violation are motivated by numerous high energy theories, including many approaches to quantum gravity [4].

If Lorentz symmetry is slightly broken in the neutrino sector, one would expect that neutrinos propagating in different directions would behave slightly differently. This could result in a change in the electron neutrino survival probability as a function of direction of propagation. Over the course of a year, the propagation direction of solar neutrinos detected at SNO rotated through a full circle, following the Earth in the frame of the Sun. SNO was therefore sensitive to such Lorentz violations as a time-of-year variation in the electron neutrino survival probability. This paper reviews the theory needed to understand how precisely to predict what would be observed in SNO and presents an analysis searching for such effects.

This paper is organized as follows. In Sec. II, we discuss the SNO detector. Section III reviews the theoretical basis of the measurement, introducing Lorentz symmetry violations in the neutrino sector in the context of the standard model extension (SME) [5,6]. In Sec. IV, we discuss some conventional effects that could give rise to similar behavior. Section V presents the analysis technique, a likelihood fit of the solar neutrino signal that includes a Lorentz violation component during the full seven-year running period of SNO. The results are presented in Sec. VI, and Sec. VII concludes.

## II. SNO DETECTOR

The Sudbury Neutrino Observatory was a heavy water Cherenkov detector, located at a depth of 2100 m (5890 m.w.e.) in Vale's Creighton mine, near Sudbury, Ontario. The detector consisted of a number of nested volumes, illustrated in Fig. 1. At the center were 1000 metric tons of $^2H_2O$ (hereafter $D_2O$ or heavy water) held in a 12-m diameter spherical acrylic vessel (AV), shown in blue in Fig. 1. Outside this was a 17.8 m diameter geodesic support structure (PSUP), which held the 9456 20-cm photomultiplier tubes (PMTs). Each PMT was fitted with a light concentrator, which increased the effective coverage of the detector to 55% [7]. The entire detector was suspended in a barrel-shaped cavity filled with ultrapure light water to act as shielding against background radiation.

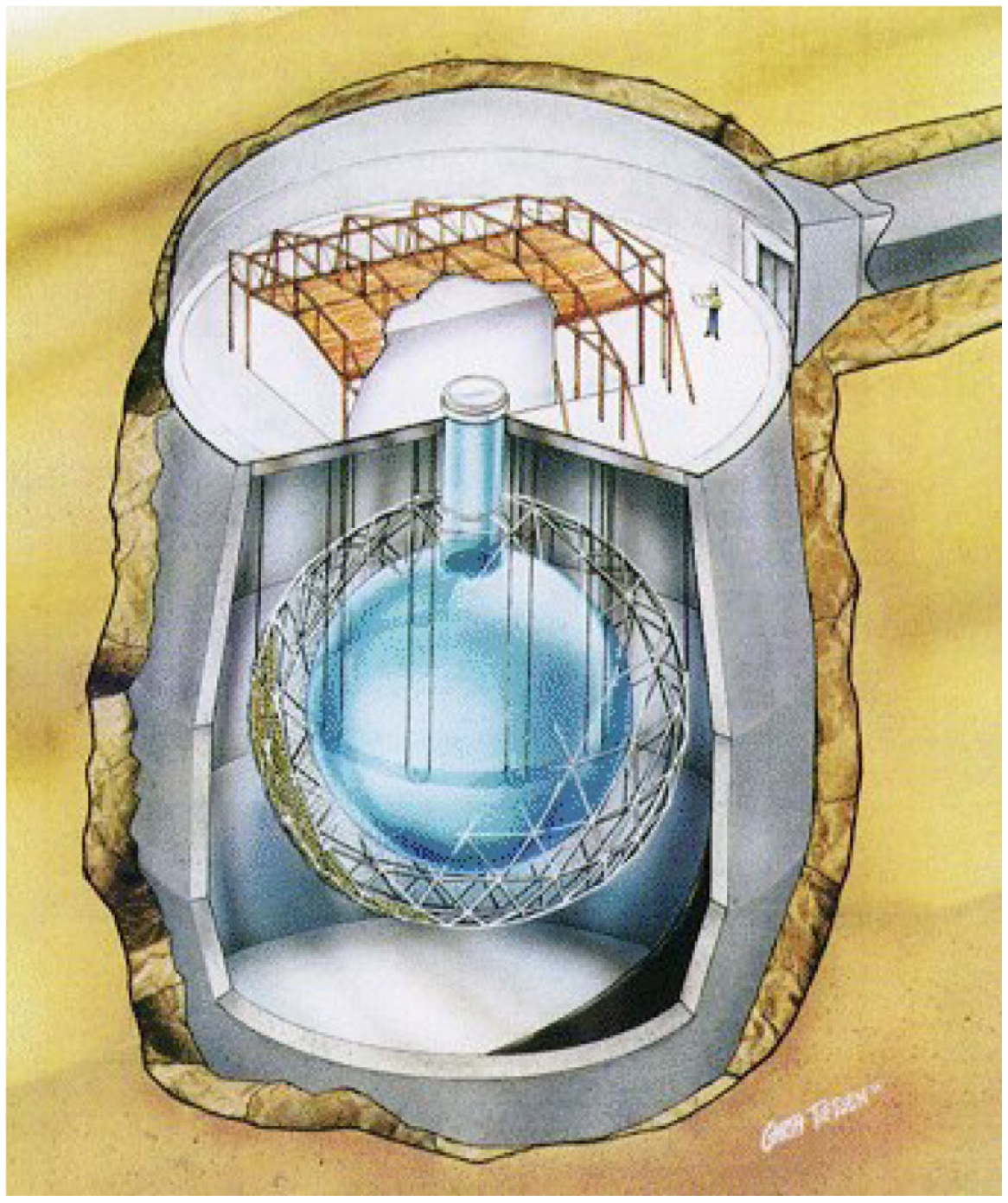

FIG. 1. The SNO detector.

SNO was sensitive to three solar neutrino interaction channels:

$$\nu_e + \mathrm{d} \to \mathrm{p} + \mathrm{p} + \mathrm{e}^- - 1.44\ \mathrm{MeV} \quad (\mathrm{CC}),$$
$$\nu + \mathrm{d} \to \mathrm{p} + \mathrm{n} + \nu - 2.22\ \mathrm{MeV} \quad (\mathrm{NC}),$$
$$\nu + \mathrm{e}^- \to \nu + \mathrm{e}^- \quad (\mathrm{ES}).$$

The neutral current (NC) interaction couples to neutrinos of all flavors equally and allowed an inclusive measurement of the active solar neutrino flux. The charged current (CC) and elastic scattering (ES) interactions couple exclusively (CC) or preferentially (ES) to the electron flavor neutrino, which allowed the solar electron neutrino survival probability to be measured.

The SNO experiment had three operational phases with different NC interaction detection techniques. In Phase I, the detector was filled with ultrapure heavy water, and the neutron liberated in the NC process was observed through its capture on deuterium. The detection rate for NC events was considerably boosted in Phase II by dissolving NaCl in the heavy water. This enabled neutron capture on chlorine, which has a higher capture cross section and produces a higher-energy signal more easily distinguished from backgrounds. In Phase III, a neutral current detection (NCD) array of $^3$He-filled proportional counters was deployed in the detector. These counters provided an independent measure of the NC event rate.

To evaluate the behavior of these signals in the SNO detector, we developed a highly detailed microphysical simulation of the detector, called SNOMAN [1]. This software was used to simulate data to reflect exactly experimental conditions at any particular time (for example, the trigger thresholds during a particular run). Samples of Monte Carlo simulations of the various signal and background events generated with statistics equivalent to many years of livetime were used extensively in this analysis.

## III. LORENTZ VIOLATION FOR SOLAR NEUTRINOS

This section provides the theoretical background for the analysis. We begin by reviewing ordinary solar neutrino oscillation before introducing the effects of possible Lorentz violations.

### A. Solar neutrino oscillation

It is well-known that there are three active neutrinos and that the weak eigenstates $|\nu_\alpha\rangle$ are mixtures of the mass eigenstates $|\nu_i\rangle$, as related by the PMNS matrix $U$, commonly parameterized in terms of mixing angles.

Neutrinos are produced in nuclear reactions in the Sun exclusively in the electron neutrino flavor. Above roughly 5 MeV, these neutrinos are then adiabatically converted nearly completely into the mass state $\nu_2$ as they pass out of the Sun due to the Mikheyev-Smirnov-Wolfenstein (MSW) or matter effect [8,9].

Vacuum oscillation effects come to dominate as the neutrino escapes the Sun, but since the neutrino energy cannot be resolved on a scale at all comparable to the number of oscillation lengths traveled from the Sun, these oscillations are averaged over. The adiabatic propagator within the Sun acts as $\hat{P}_1|\nu_i\rangle = e^{i\phi_i}|\nu_i\rangle$. The vacuum oscillation propagator is $\hat{P}_2 = e^{-im^2L/2E}$ (in the ultrarelativistic limit). Thus, the oscillation probability can be computed by

$$\begin{aligned} P_{\beta\alpha} &= |\langle\nu_\alpha|\hat{P}_2\,\hat{P}_1\,|\nu_\beta\rangle|^2 \\ &= |\langle\nu_\alpha|e^{-im^2L/2E}\hat{P}_1|\nu_\beta\rangle|^2 \\ &= \left|\sum\nolimits_{ij}\langle\nu_i|U^*_{\alpha i}e^{-im^2L/2E}\hat{U}_{\beta j}\hat{P}_1|\nu_j\rangle\right|^2. \end{aligned} \quad (1)$$

Here $\hat{U}$ represents the matter-perturbed mixing matrix relevant for a solar neutrino at its creation. This depends on both its radial position within the Sun (since it is electron-density dependent) as well as the energy of the neutrino in question.

Applying the adiabatic propagator, we find

$$\begin{aligned} P_{\beta\alpha} &= \left|\sum_{ij}U^*_{\alpha i}e^{-im_i^2L/2E}e^{i\phi_i}\delta_{ij}\hat{U}_{\beta j}\right|^2 \\ &= \sum_{ij}U^*_{\alpha i}U_{\alpha j}\hat{U}_{\beta i}\hat{U}^*_{\beta j}e^{i\Delta m^2_{ji}L/2E+i\phi_{ij}} \\ &= \sum_i|U_{\alpha i}\hat{U}_{\beta i}|^2. \end{aligned} \quad (2)$$

The last step follows because the phase will average to zero unless $i = j$ (again, since the phase will vary enormously between neutrinos of very similar energies). Here $\Delta m^2_{ij} \equiv m_i^2 - m_j^2$ and $\phi_{ij} \equiv \phi_i - \phi_j$.

### B. Lorentz violation in the neutrino sector

A consistent framework for discussing violations of Lorentz symmetry was introduced by Kostelecký *et al.* [5,6], in what is called the standard model extension (SME). This framework includes all possible Lorentz violating operators of the standard model particle fields while retaining causality and observer independence. (The theory is invariant under boosts to observers, but not under boosts to particles.) The SME includes a large number of such operators, each controlled by a distinct coefficient determining the size of that effect. This framework has been widely adopted by experimentalists in reporting the limits established in a variety of areas [10].

An explication of the SME framework in the neutrino sector is given in [11]. Here we extend the discussion of the prediction of the model for solar neutrinos given in [12] to operators of arbitrary dimension and update the discussion to use the spherical harmonic decomposition introduced [11] after it was written.

We assume that the neutrino Hamiltonian is dominated by the usual mass and matter terms, with effects due to Lorentz violation forming a small perturbation to the usual dynamics [11]:

$$\delta H = \frac{1}{|p|}\begin{pmatrix} a_{\rm eff} - c_{\rm eff} & -g_{\rm eff} + H_{\rm eff} \\ -g^\dagger_{\rm eff} + H^\dagger_{\rm eff} & -a^T_{\rm eff} - c^T_{\rm eff} \end{pmatrix}. \quad (3)$$

This expression is written in a block matrix form, with $a_{\rm eff}$, $c_{\rm eff}$, $g_{\rm eff}$, and $H_{\rm eff}$ standing for $3\times3$ matrices that determine the size of Lorentz violating effects, and $p$ is the neutrino momentum. The upper three coordinates are for the three flavors of neutrinos, and the lower three coordinates are for the antineutrinos.

Because it is expected that $a$- and $c$-type terms and $g$- and $H$-type terms would arise from different underlying physics, and because SNO is sensitive to these terms at widely different levels, it is reasonable to perform an analysis for each of these types of terms separately. In this analysis, we consider only $a$ and $c$ terms and assume that $g$ and $H$ are negligibly small. We therefore focus exclusively on the upper left (neutrino-neutrino) quadrant of the Hamiltonian.

Each of these terms can be expanded in a series of operators of increasing mass dimension, $d$ [11]. For example,

$$a^{ab}_{\rm eff} = \sum_{djm}|p|^{d-2}Y_{jm}(\hat{p})(a^{(d)}_{\rm eff})^{ab}_{jm}, \quad (4)$$

where $Y_{jm}$ are the spherical harmonic functions. This representation makes it clear that there are in principle an infinite number of possible effects to consider. We therefore need a criterion for selecting a subset for which to search. One particularly straightforward choice is to consider only renormalizable terms. As the lowest-dimension operators, these are likely to be the most important. Some of the renormalizable terms are helicity-suppressed at leading order, and we do not search for such effects in this analysis. The remaining terms (the dominant terms in the minimal SME) are three spherical-harmonic multipoles of $c^{(4)}_{\rm eff}$ and two multipoles of $a^{(3)}_{\rm eff}$ [11]. The correction to the neutrino Hamiltonian we use is, therefore,

$$\delta H = \sum_{jm} Y_{jm}(\hat{p})(a^{(3)}_{\rm eff})_{jm} - \sum_{jm} |p| Y_{jm}(\hat{p})(c^{(4)}_{\rm eff})_{jm}, \quad (5)$$

where $j$ is 0 or 1 for the $a$ term and 0, 1, or 2 for the $c$ term. In principle, $m$ can take integer values between $-j$ and $j$. However, there are relationships between different coefficients which arise from the fact that the coefficients are Hermitian in flavor space [11]. This reduces the total number of independent degrees of freedom. For $a$ and $c$ coefficients, results are typically presented in terms of the real and imaginary parts of the coefficients with nonnegative $m$. Since the $m = 0$ term is real, this results in a total of $2j + 1$ degrees of freedom.

In this analysis, we search for these effects individually, assuming that all others are zero, as is usual in searches of this kind (e.g., [13,14]). Were Lorentz violations observed in a system, a more sophisticated analysis fitting for multiple types of violations simultaneously would be desirable.

### C. Lorentz violation in solar neutrinos

Since we have assumed that $\delta H$ is small in comparison to the conventional mass and matter terms, we can apply perturbation theory and assume that the full Hamiltonian will be diagonalized by a matrix $U$, with $U = U^{(0)} + \delta U$. We define for convenience

$$U = (1 + \xi)U^{(0)}, \quad (6)$$

so that

$$\delta U = \xi U^{(0)}. \quad (7)$$

Then recognizing that $(I + \xi)$ diagonalizes $U^{(0)}\delta H U^{(0)\dagger}$, we can use ordinary perturbation theory to conclude that to first order the matrix elements of $\xi$ are given by the corrections to the eigenstates of $H^{(0)}$. For the off-diagonal elements,

$$\xi_{kj} = \frac{(U^{(0)}\delta H U^{(0)\dagger})_{kj}}{E_j - E_k} = \sum_{\alpha\beta} \frac{U^{(0)}_{j\alpha} U^{(0)*}_{k\beta}}{E_j - E_k} \delta H_{\alpha\beta}, \quad (8)$$

where $E_i$ is the energy of the $i$th unperturbed mass state. The diagonal elements of $\xi$ are identically zero at first order.

Combining this expression with Eq. (7), we see

$$\delta U_{i\gamma} = \sum_{\alpha\beta j} \frac{U^{(0)}_{j\alpha} U^{(0)*}_{i\beta} U^{(0)}_{j\gamma}}{E_j - E_i} \delta H_{\alpha\beta}. \quad (9)$$

We can then compute the first-order corrections to the transition probabilities following Eq. (2):

$$\begin{aligned} P_{\beta\alpha} &= \sum_i |(U^{(0)} + \delta U)_{i\alpha} (\hat{U}^{(0)} + \delta\hat{U})_{i\beta}|^2 \\ &= P^{(0)}_{\beta\alpha} + 2\mathfrak{Re} \sum_i U^{(0)}_{i\alpha} \hat{U}^{(0)}_{i\beta} (U^{(0)*}_{i\alpha} \delta\hat{U}^*_{i\beta} + \delta U^*_{i\alpha} \hat{U}^{(0)*}_{i\beta}). \end{aligned} \quad (10)$$

Plugging in our expression for $\delta U$, we then obtain the correction to the probability:

$$\begin{aligned} \delta P^{(1)}_{\beta\alpha} &= 2\sum_{\gamma\delta kl} |U^{(0)}_{k\alpha}|^2 \mathfrak{Re}\left(\hat{U}^{(0)}_{k\beta} \frac{\hat{U}^{(0)*}_{l\gamma} \hat{U}^{(0)}_{k\delta} \hat{U}^{(0)*}_{l\beta}}{\hat{E}_l - \hat{E}_k} \delta H_{\gamma\delta}\right) \\ &\quad + |\hat{U}^{(0)}_{k\beta}|^2 \mathfrak{Re}\left(U^{(0)}_{k\alpha} \frac{U^{(0)*}_{l\gamma} U^{(0)}_{k\delta} U^{(0)*}_{l\alpha}}{E_l - E_k} \delta H_{\gamma\delta}\right). \end{aligned} \quad (11)$$

Finally, we substitute Eq. (5) to get our final expression for the changes to the oscillation probabilities:

$$\begin{aligned} \delta P^{(1)}_{\beta\alpha} &= 2\sum_{jm} \mathfrak{Re}\left\{ Y_{jm}(\hat{p}) \sum_{\gamma\delta} ((a^{(3)}_{\rm eff})^{\gamma\delta}_{jm} - E(c^{(4)}_{\rm eff})^{\gamma\delta}_{jm}) \right. \\ &\quad \times \sum_{kl} \left( |U^{(0)}_{k\alpha}|^2 \hat{U}^{(0)}_{k\beta} \frac{\hat{U}^{(0)*}_{l\gamma} \hat{U}^{(0)}_{k\delta} \hat{U}^{(0)*}_{l\beta}}{\hat{E}_l - \hat{E}_k} \right. \\ &\quad \left.\left. + |\hat{U}^{(0)}_{k\beta}|^2 U^{(0)}_{k\alpha} \frac{U^{(0)*}_{l\gamma} U^{(0)}_{k\delta} U^{(0)*}_{l\alpha}}{E_l - E_k} \right) \right\}. \end{aligned} \quad (12)$$

In Eq. (12), the final sum over $kl$ is just a function of energy that depends on the neutrino mixing angles, masses, and the matter potential present in the Sun. However, it is independent of the size of the Lorentz-violating effects and the propagation direction, which provides a useful simplification. This function specifies the (energy-dependent) linear combination of Lorentz-violating fields to which the experiment is sensitive. After factoring out the dominant linear energy dependence, we denote this by $w$:

TABLE I. Mixing model parameter values used in the analysis. PDG values without input from SNO [15].

| Mixing model parameter | Value |
|---|---|
| $\sin^2\theta_{13}/10^{-2}$ | $2.10 \pm 0.11$ |
| $\Delta m^2_{12}/10^{-5}$ eV$^2$ | $7.54 \pm 0.19$ |
| $\Delta m^2_{23}/10^{-3}$ eV$^2$ | $2.48 \pm 0.08$ |

$$w^{\beta\alpha}_{\gamma\delta} = \frac{2}{E}\sum_{ij}\left(|U^{(0)}_{i\alpha}|^2\left(\hat{U}^{(0)}_{i\beta}\frac{\hat{U}^{(0)}_{j\gamma}\hat{U}^{(0)*}_{i\delta}\hat{U}^{(0)}_{j\beta}}{\hat{E}_j-\hat{E}_i}\right)\right.$$
$$\left.+|\hat{U}^{(0)}_{i\beta}|^2\left(U^{(0)}_{i\alpha}\frac{U^{(0)}_{j\gamma}U^{(0)*}_{i\delta}U^{(0)}_{j\alpha}}{E_j-E_i}\right)\right). \quad (13)$$

The survival probability correction can then be written compactly as:

$$\delta P^{(1)}_{\beta\alpha} = \mathfrak{Re}\sum_{jm\gamma\delta} Y_{jm}(\hat{p}) w^{\beta\alpha}_{\gamma\delta}(E(a^{(3)}_{\text{eff}})^{\gamma\delta}_{jm} - E^2(c^{(4)}_{\text{eff}})^{\gamma\delta}_{jm}). \quad (14)$$

We calculated $w_{\gamma\delta}$ for electron neutrino survival over the energy range relevant for solar neutrinos, namely 1–20 MeV, using the model parameters defined in Table I. These weight functions are shown in Fig. 2. It can be seen that the different contributions become relatively constant at energies above about 6 MeV, after the MSW transition has saturated.

### D. Independent observables in SNO

For each distinct energy and time behavior (choice of $d$, $j$, and $m$), there is a group of nine nearly-degenerate Lorentz violating coefficients differing only by their weight function. These have slightly different energy dependencies at lower energies, where the mixing angles change significantly as a result of the MSW effect in the Sun.

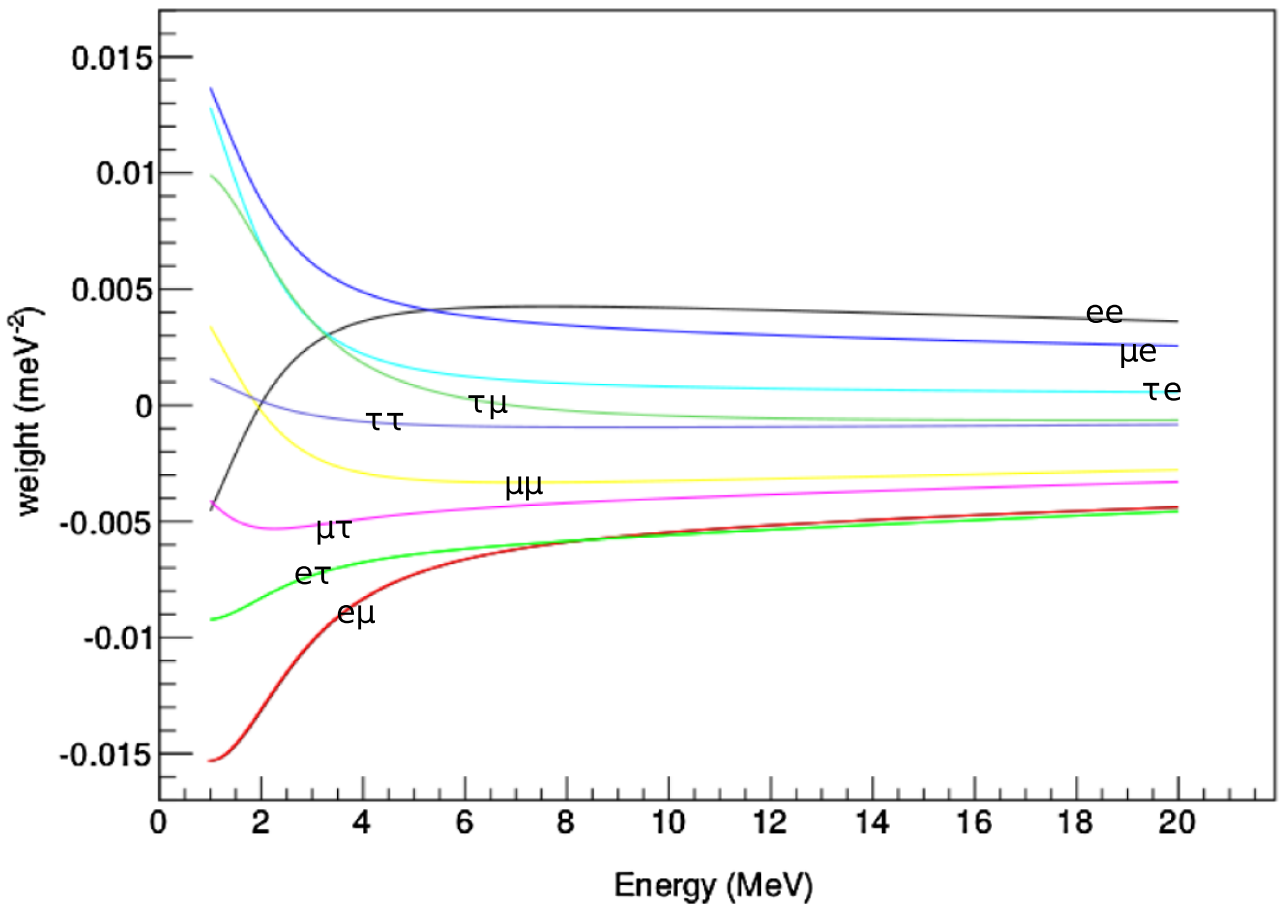

FIG. 2. Weight functions $w^{ee}_{\gamma\delta}$ for the various coefficients as a function of energy. Above 6 MeV, the contributions are reasonably constant. Each color represents a different flavor pair $\gamma\delta$, as labeled.

There are two possible approaches to handling this near degeneracy. Either we restrict ourselves to a domain in which the signals are truly degenerate, compute the linear combination of coefficients to which we are sensitive, and set a single limit, or we keep all signals distinct and try to fit for them simultaneously. Even at the lowest energy threshold used by SNO (3.5 MeV), the shapes of the different effects are not obviously resolved. An unrealistically optimistic sensitivity study showed that the global correlation of each of the Lorentz-violating parameters was at least 0.985. This confirmed that there is no power to distinguish the different effects. It is therefore necessary for the analysis to take the other approach, namely, to search for the single linear combination of these effects to which the detector is sensitive. For this analysis, we apply a lower energy threshold of 7 MeV. This puts us firmly in the regime where the weights are independent of energy and also reduces the risk of contamination from radioactive backgrounds.

We define the SNO combination of effects as

$$c^{(4)}_{\text{SNO}} = \sum_{\alpha\beta} w^{ee}_{\alpha\beta}(c^{(4)}_{\text{eff}})^{\alpha\beta}, \quad (15)$$

with $a^{(3)}_{\text{SNO}}$ defined analogously. With these definitions, the probability simplifies to

$$\delta P^{(1)}_{ee} = \mathfrak{Re}\sum_{jm} Y_{jm}(\hat{p})(E(a^{(3)}_{\text{SNO}})_{jm} - E^2(c^{(4)}_{\text{SNO}})_{jm}). \quad (16)$$

It is this expression that is used in practice for fitting the data.

To zeroth order, $a_{\text{SNO}}$ and $c_{\text{SNO}}$ can be read off the plot in Fig. 2, but a more detailed treatment, taking into account the standard solar model, will be discussed below. For setting limits, the SNO weight combination is computed using the fit result for each mode separately. The final results are reported in Sec. VI.

After grouping the nearly degenerate parameters into effective parameters, there are still four parameters of dimension three and nine of dimension four. Those of dimension three produce signals that cycle at most once per year and grow linearly with energy. Those of dimension four grow quadratically with energy (and are therefore independent of the dimension-three operators) and have signals that cycle at most twice per year. Simple considerations from Fourier analysis show that there can be at most three and five independent observables in these two cases. We therefore decompose the signals into their Fourier modes and summarize these combinations in Table II.

TABLE II. Table of independent observables, and the terms in the theory that contribute to each. $\Omega$ is the orbital inclination of the Earth, and $\omega$ is the orbital frequency of the Earth.

| Signal | Source (algebraic) | Source (numeric) |
|---|---|---|
| $E$ | $\frac{1}{2}\sqrt{\frac{1}{\pi}}(a^{(3)}_{\mathrm{SNO}})_{00}$ | $0.28a_{00}$ |
| $E\sin\omega t$ | $\frac{1}{2}\sqrt{\frac{3}{\pi}}\sin\Omega(a^{(3)}_{\mathrm{SNO}})_{10} - \frac{1}{2}\sqrt{\frac{3}{2\pi}}\cos\Omega\mathfrak{Im}(a^{(3)}_{\mathrm{SNO}})_{11}$ | $0.19a_{10} - 0.32\mathfrak{Im}a_{11}$ |
| $E\cos\omega t$ | $\frac{-1}{2}\sqrt{\frac{3}{2\pi}}\mathfrak{Re}(a^{(3)}_{\mathrm{SNO}})_{11}$ | $-0.35\mathfrak{Re}a_{11}$ |
| $E^2$ | $\frac{1}{2}\sqrt{\frac{1}{\pi}}(c^{(4)}_{\mathrm{SNO}})_{00} + \frac{1}{4}\sqrt{\frac{15}{4\pi}}(1 - \frac{1}{2}\sin^2\Omega - \cos^2\Omega)\mathfrak{Re}(c^{(4)}_{\mathrm{SNO}})_{22}$ $+\frac{1}{4}\sqrt{\frac{5}{\pi}}(\frac{3}{2}\sin^2\Omega - 1)(c^{(4)}_{\mathrm{SNO}})_{20} - \frac{1}{4}\sqrt{\frac{15}{2\pi}}\sin\Omega\cos\Omega\mathfrak{Im}(c^{(4)}_{\mathrm{SNO}})_{21}$ | $0.28c_{00} + 0.03\mathfrak{Re}c_{22} - 0.08c_{20} - 0.14\mathfrak{Im}c_{21}$ |
| $E^2\sin\omega t$ | $\frac{1}{2}\sqrt{\frac{3}{\pi}}\sin\Omega(c^{(4)}_{\mathrm{SNO}})_{10} - \frac{1}{2}\sqrt{\frac{3}{2\pi}}\cos\Omega\mathfrak{Im}(c^{(4)}_{\mathrm{SNO}})_{11}$ | $0.19c_{10} - 0.32\mathfrak{Im}c_{11}$ |
| $E^2\cos\omega t$ | $\frac{-1}{2}\sqrt{\frac{3}{2\pi}}\mathfrak{Re}(c^{(4)}_{\mathrm{SNO}})_{11}$ | $-0.35\mathfrak{Re}c_{11}$ |
| $E^2\sin 2\omega t$ | $\frac{-1}{4}\sqrt{\frac{15}{2\pi}}\sin\Omega\mathfrak{Re}(c^{(4)}_{\mathrm{SNO}})_{21} + \frac{1}{4}\sqrt{\frac{15}{2\pi}}\cos\Omega\mathfrak{Im}(c^{(4)}_{\mathrm{SNO}})_{22}$ | $-0.15\mathfrak{Re}c_{21} + 0.35\mathfrak{Im}c_{22}$ |
| $E^2\cos 2\omega t$ | $\frac{-3}{8}\sqrt{\frac{5}{\pi}}\sin^2\Omega(c^{(4)}_{\mathrm{SNO}})_{20} + \frac{1}{4}\sqrt{\frac{15}{2\pi}}\sin\Omega\cos\Omega\mathfrak{Im}(c^{(4)}_{\mathrm{SNO}})_{21}$ $+\frac{1}{4}\sqrt{\frac{15}{2\pi}}(\frac{1}{2}\sin^2\Omega + \cos^2\Omega)\mathfrak{Re}(c^{(4)}_{\mathrm{SNO}})_{22}$ | $-0.08c_{20} + 0.14\mathfrak{Im}c_{21} + 0.36\mathfrak{Re}c_{22}$ |

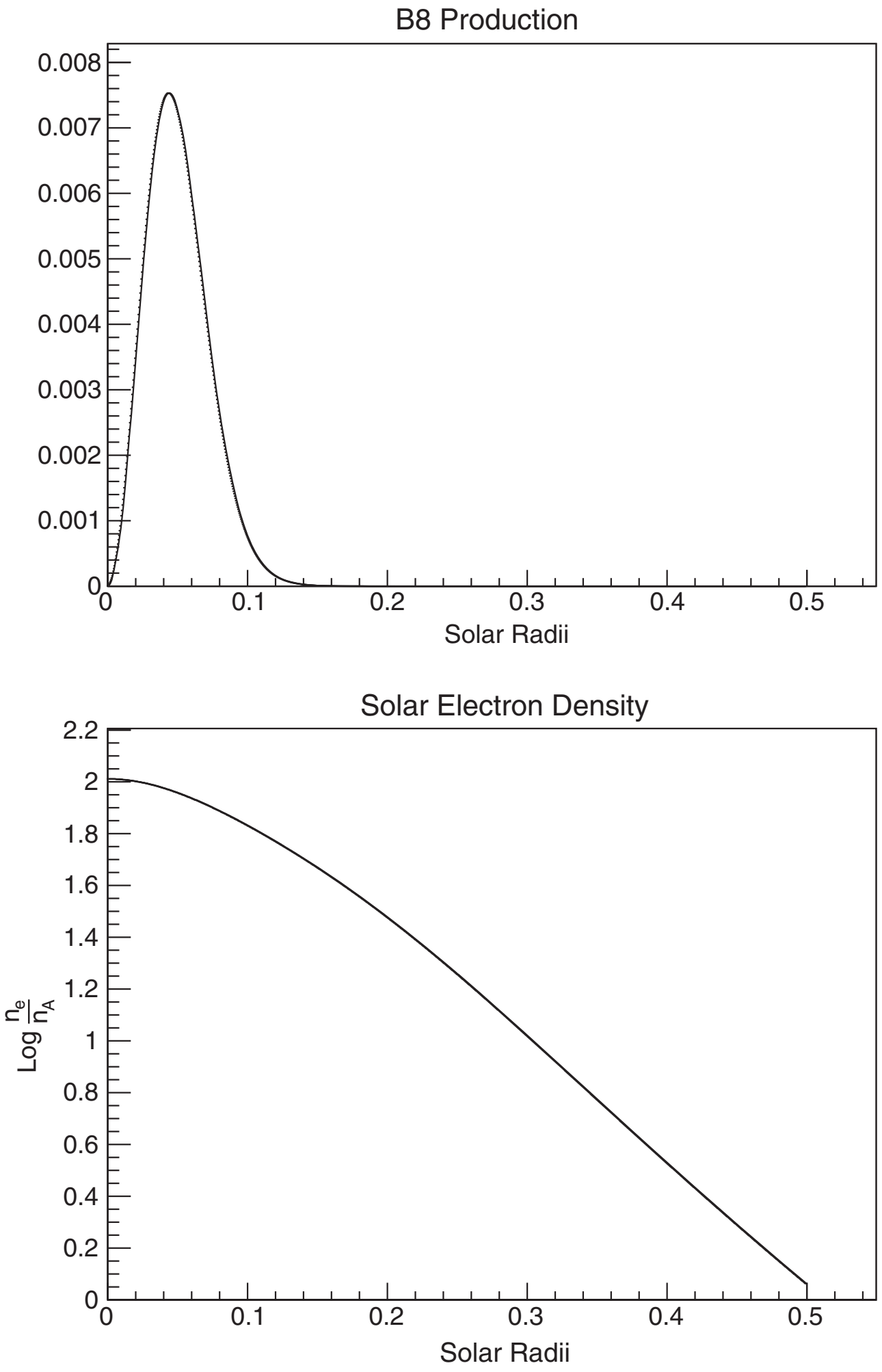

FIG. 3. Above: Distribution of $^8$B production radii. Below: Solar electron number density as a function of radius, plotted as Log (base 10) of $n_e$ per cm$^3$ per $n_A$. Data taken from [16].

### E. Modeling the signal

SNO was sensitive to neutrino flavor through the ES and CC interactions. There is no change to the flavor-blind NC interaction from Lorentz violations, since this is not affected by the electron neutrino survival probability (since we are assuming $g$ and $H$ are zero).

To provide intuition about what the signal would look like in the SNO detector, we propagate the changes to the survival probability through the nuclear interactions and detector effects by reweighting the SNO Monte Carlo data.

The Sun is not homogeneous, so neutrinos coming from different locations within the Sun will behave slightly differently. We model this according to the standard solar model (BS05(OP)) [16]. As can be seen from Eq. (16), this only has an effect on the particular linear combination of coefficients to which we are sensitive and not on the shape of the signal. The data on the radial distribution of $^8$B production and the electron number density in the Sun, taken from [16], are shown in Fig. 3.

For each SNO $^8$B Monte Carlo event, we randomly sampled a solar origin point and used this to calculate its survival probability for any choice of the mixing parameters. Templates for the changes to the reconstructed $^8$B CC and ES energy spectrum for a fixed value of the mixing parameters (as defined in Table I) and for a Lorentz-violating coefficient $c^{(4)}_{\mathrm{SNO}} = 10$ GeV$^{-2}$ are shown in Fig. 4.

These templates are meant only for illustration, and are not used explicitly in the final fit. The fit instead uses probability density functions (PDFs) for solar events which include both the standard model and Lorentz violating effects together. In the fit, the details of the standard solar model are included when the SNO combinations are computed, as reported in Sec. VI, since this is significantly more computationally efficient than including the effect in the PDFs.

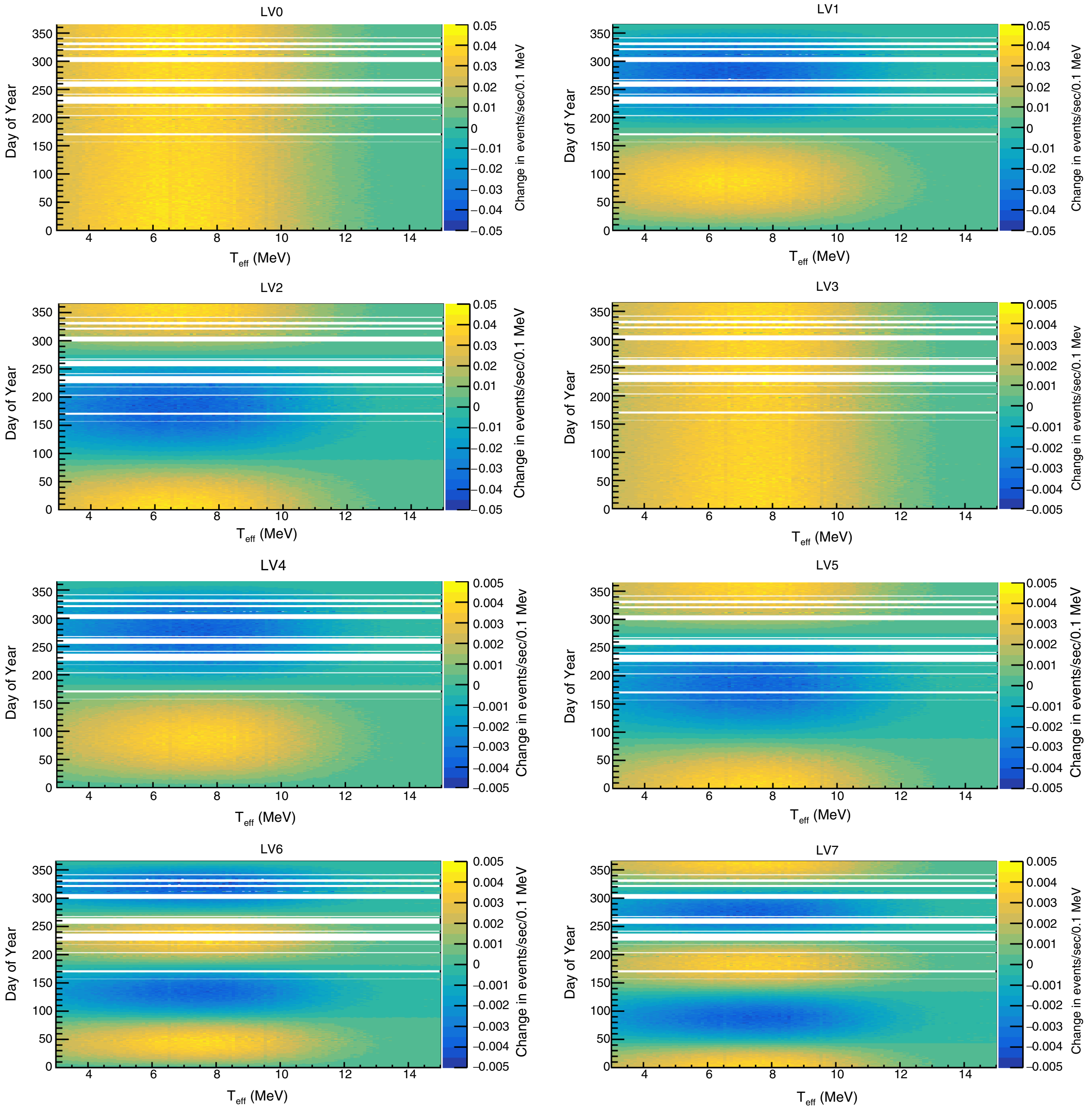

FIG. 4. The eight signal types expected in SNO if $a^{(3)}_{\mathrm{SNO}} = 0.01\ \mathrm{GeV}^{-1}$ and $c^{(4)}_{\mathrm{SNO}} = 10\ \mathrm{GeV}^{-2}$. The figure was generated by taking the $^{8}$B CC and ES Monte Carlo events for the selected runs in Phases I and II and weighing each event by the correction to the survival probability, then normalizing by the livetime. White areas denote days for which there was no livetime. LV0, LV1, and LV2 are proportional to $E$, while the other signals are proportional to $E^2$, and can therefore be seen to be shifted slightly toward higher energies.

## IV. COMPETING EFFECTS

We investigated two known effects that induce seasonal variations in the solar neutrino flux, since such behavior could confound the analysis and must be controlled for. The first of these is the eccentricity of the Earth's orbit ($\epsilon = 0.0167$ [17]), which leads to a 3% annual variation in the neutrino flux. We compute the Earth's Keplerian orbit and explicitly include these effects on the flux and the Sun-to-Earth direction in our model.

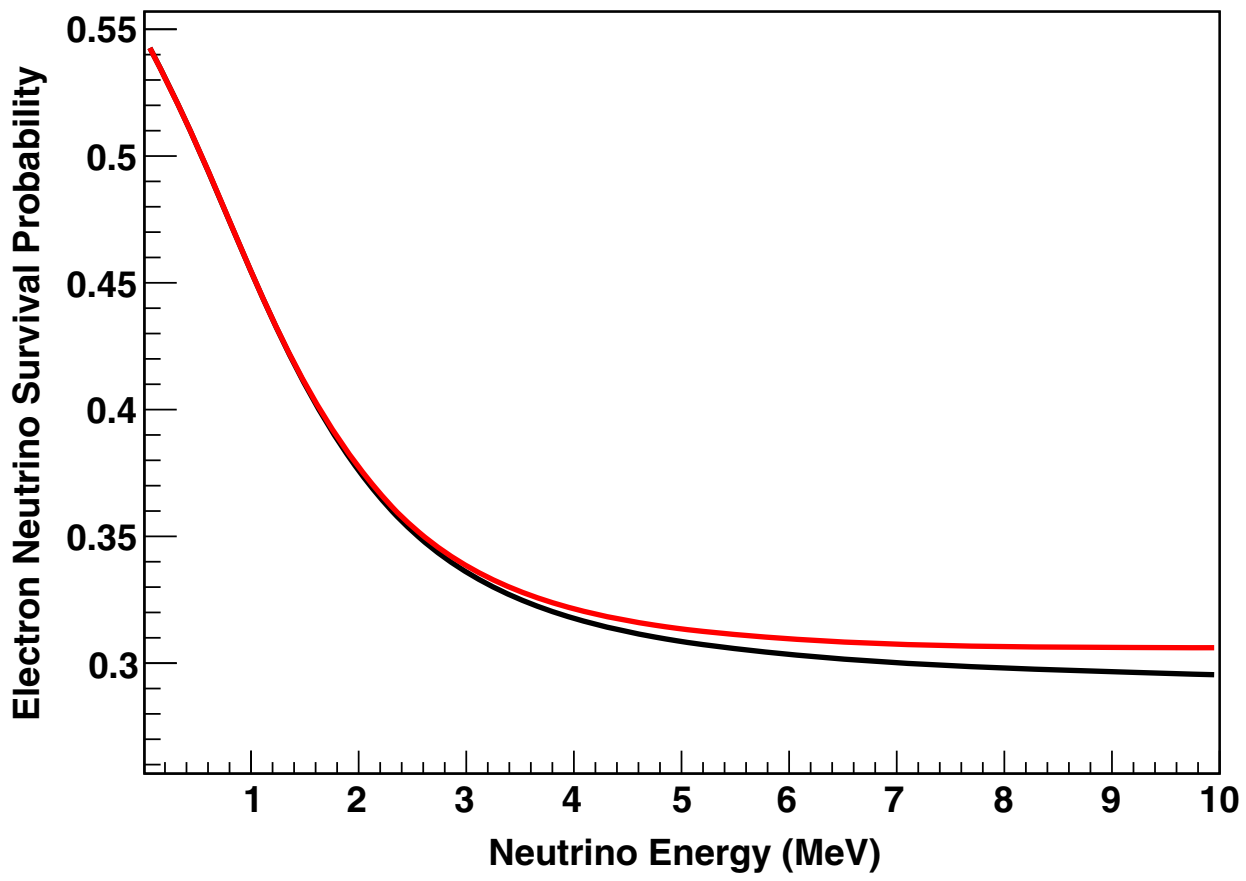

FIG. 5. Survival probability in day (black) and night (red). Effect calculated assuming $\sin\theta_{12} = 0.555$.

A second competing effect is caused by a difference in the $\nu_e$ survival probability between day and night due to the regeneration of $\nu_e$ from matter effects in the Earth. Because the fraction of day and night varies systematically over the course of the year, it is important to control for this effect. The effect in a SNO-like detector, averaged over a year, was derived in [18]. This effect depends only on the local electron number density in the vicinity of the detector and can be expressed as $p_{\text{night}} = p_{\text{day}} \frac{1+\delta}{1-\delta}$ with

$$\delta = \frac{-\cos 2\theta_m \sin^2 2\theta}{1 + \cos 2\theta_m \cos 2\theta} \frac{EV}{\Delta m^2} \tag{17}$$

where $V$ is the matter potential in the vicinity of the detector. In [18] it is shown that $V = (1.1 \pm 0.1) \times 10^{-10}$ meV more than covers any errors between the exact result and this approximation. The shape of this effect is illustrated in Fig. 5.

In addition to this year-averaged effect, the nighttime survival probability also varies seasonally as different parts of the Earth's interior are probed. Reference [19] gives an expression for the instantaneous nighttime survival probability in an adiabatic approximation. We applied this analytic solution to a simplified version of the Preliminary Reference Earth Model [20], consisting of four layers of constant density, to estimate the size of this seasonal effect. The magnitude of the seasonal effect is strongly enhanced at stationary points (at midnight and at the solstices). Figure 6 shows the nighttime survival probability at midnight on each night of the year as a function of neutrino energy. Our model was verified for consistency with the results in [19] and found to agree at the level of 10%.

Although the true effect on the neutrino survival probability is quite substantial, the observed effect is washed out considerably because of: (i) the kinematic smearing inherent in the CC and ES processes; (ii) the intrinsic energy resolution of the detector; and (iii) reductions of day-to-day variations when averaged over seasonal time periods.

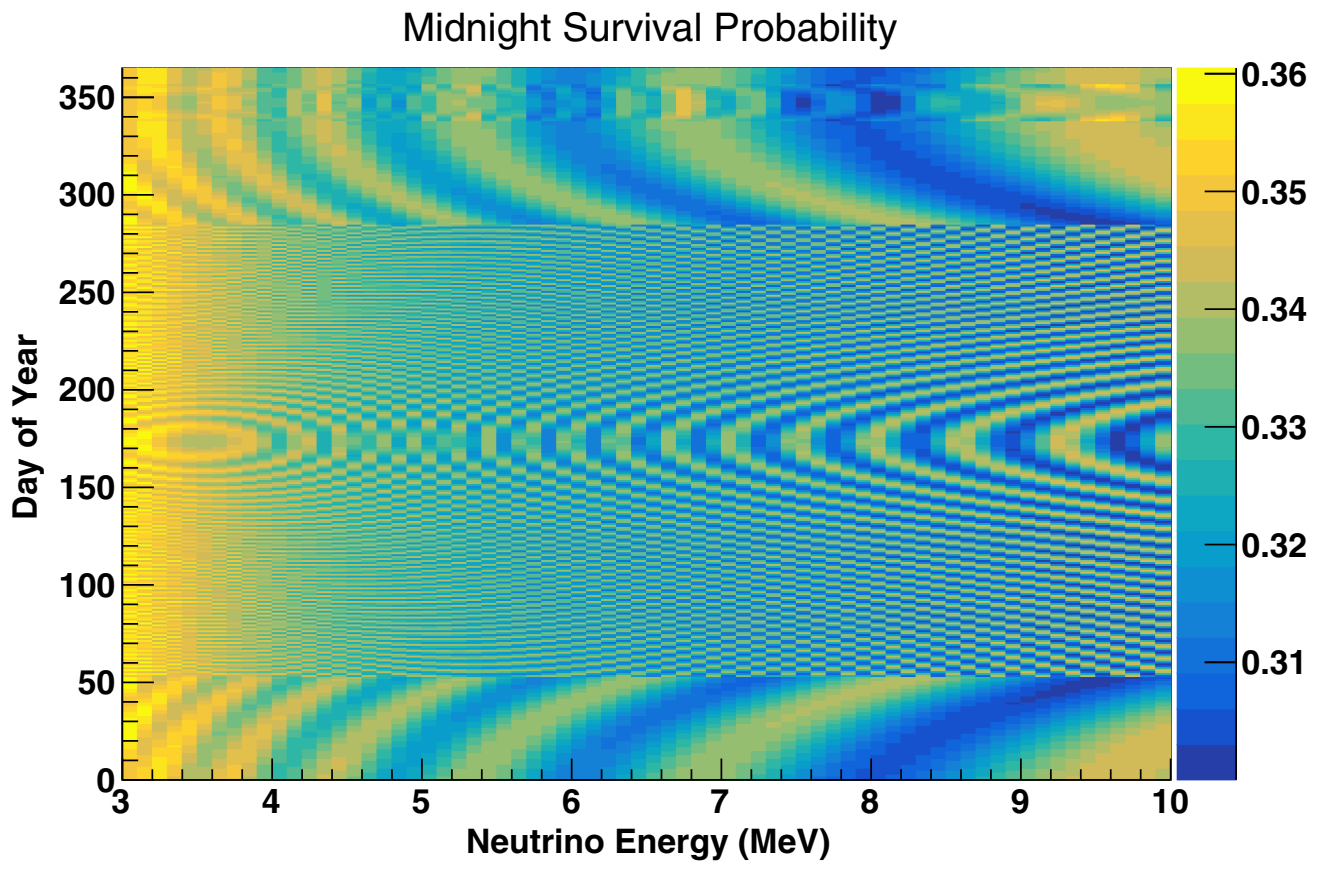

FIG. 6. Seasonal variation in the nighttime survival probability due to changes in electron neutrino regeneration in the Earth. During the broad summertime period, neutrinos traverse only the outermost layers of the Earth and, except during the summer solstice, the effect averages away over short timescales. In the winter, contributions from the denser parts of the Earth become relevant and the time and energy scales of the effect become more easily resolved by a SNO-like detector. The period around the winter solstice is the only time during which neutrinos traverse the (outer) core.

Figure 7 shows the result of convolving this effect with the detector response. The magnitude of the effect is reduced roughly by a factor of 5.

Because this effect has a very detailed structure which is not robust to small changes in the model (e.g., the Earth density profile), its inclusion in the model used to fit the data would be liable to introduce an error as large as the one it intends to remove. We therefore treat this as a systematic uncertainty. The bias introduced to the Lorentz violation parameter is negligibly small (3–9% of the statistical uncertainty). It also introduces a small bias to the mixing parameters (about 3% for the solar flux and 1% for the survival probability). This is because the actual data are not sampled exactly uniformly across the year, so using the year-averaged nighttime probability is not necessarily an accurate estimate of the nighttime probability observed in the data. This effect is also quite modest and is not of direct interest to this analysis.

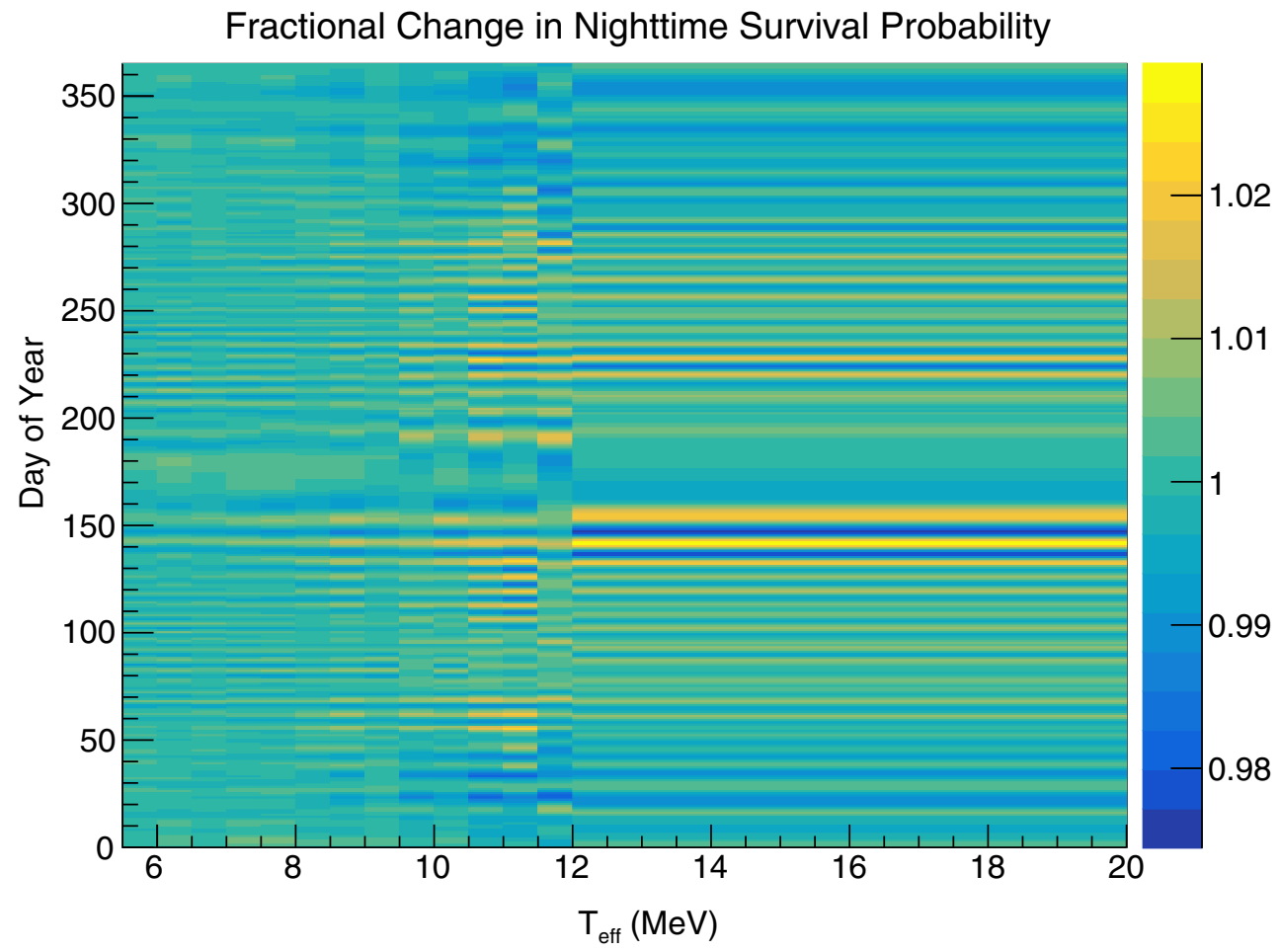

FIG. 7. Seasonal day-night effect after convolving with detector response. The size of the effect is reduced considerably. The figure shows an average of Phases I and II.

## V. ANALYSIS

We performed a separate likelihood search for each of the eight possible signal types defined in Table II. Each fit has three free parameters: $\sin\theta_{12}$, the solar $^8$B flux, and one Lorentz symmetry violating parameter. The other mixing parameters are fixed at the levels shown in Table I, with uncertainties handled as systematics.

### A. Data selection

Data selection proceeds in a number of steps. The data are organized in typically hours-long periods called runs, and the first step is to select for runs with stable detector conditions. This analysis uses the same run list developed for the full analysis of all three phases of the SNO data [21].

There is also an event-level selection within each run. These cuts are designed to remove instrumental backgrounds and eliminate muons and muon-induced backgrounds from the data set. Again, for this analysis we use the same reconstruction corrections, and data cleaning and high-level cuts used in [21] for identifying physics events.

We define a region of interest for the analysis in terms of effective recoil electron kinetic energy $T_{\rm eff}$ and radial position $r$, requiring $r < 5.5$ m, and $7.0\ \mathrm{MeV} < T_{\rm eff} < 20$ MeV. The low-energy threshold for this analysis was selected to optimize the sensitivity of the measurement. The interplay between the loss of signal and growing systematic uncertainties is summarized in Table III. This threshold differs signficantly from that used in [21] because of the different systematic concerns germane to this analysis, most prominently the time-stability of background levels. We discuss these systematic uncertainties in detail in Sec. V F.

TABLE III. Contributions to sensitivity to $a^{(3)}_{\rm SNO}$ in units of GeV$^{-1}$ as a function of energy threshold. The sensitivity to $c^{(4)}_{\rm SNO}$ is proportional.

| Threshold | Systematic | Statistical | Total |
|---|---|---|---|
| 5.5 MeV | 0.192 | 0.364 | 0.411 |
| 6.0 MeV | 0.158 | 0.378 | 0.410 |
| 6.5 MeV | 0.129 | 0.387 | 0.408 |
| 7.0 MeV | 0.101 | 0.395 | 0.408 |
| 7.5 MeV | 0.076 | 0.412 | 0.419 |
| 8.0 MeV | 0.056 | 0.422 | 0.426 |

### B. Blindness

The data remained blinded during the development of the analysis by removing events that reconstructed in the region of interest from the data set. In this development period, we studied sideband regions or used Monte Carlo simulations of the blinded region. Once the analysis was finalized, the data were unblinded in two stages. The fit was first run on a one-third statistical subsample to verify that it behaved as expected on real data before proceeding to fit the full data set.

### C. Fit

We developed a binned likelihood fit that consists of three pieces. (i) For Phases I and II, we perform a fit in energy ($T_{\rm eff}$), volume-weighted radius ($\rho = r^3/r^3_{AV}$), solar angle, and isotropy ($\beta_{14}$). (ii) For Phase III PMT data, we perform a fit in energy, radius, and solar angle. For each of these components, the binning of the observables used were those in [21], but with the lowest-energy bins excluded. (iii) For Phase III NCD data, we use a constraint from the earlier pulse shape analysis [22] that determined that $1115 \pm 79$ NCD events were due to physics events. This was fit with a model of signal and background NCD interactions also used in [21].

We considered the impact of many possible systematic effects on the analysis, including uncertainties in the shape and normalization of the PDFs used in fit, time variations in background event rates, and uncertainties in the neutrino mixing model. Because we found that the measurement was ultimately statistically limited, the systematic error was estimated using a shift-and-refit strategy. These systematics are discussed in detail in Sec. V F.

### D. Backgrounds

Besides instrumental backgrounds which can be easily removed with data cleaning cuts based on event topology, the two main sources of background physics events are radioactive backgrounds and atmospheric neutrino interactions. In spite of the very successful efforts to reduce radioactivity levels in the detector, some residual U/Th chain contamination remains. Decays of this material lead directly to $\beta$'s or $\gamma$'s that can Compton scatter in the detector. These decays are classified as "internal" if they actually occur within the region of interest or as "external" if they occur outside the AV but either scatter into the region of interest or misreconstruct there.

At higher energies, radioactive backgrounds dwindle in number and importance. The number of events in the selection used here not caused by $^8$B solar neutrinos is estimated to be about 2%, with Phases II and III having higher background rates than Phase I because of the additional materials and poorer energy resolution of those phases. Estimates of the contributions within the analysis region from signal and background events are shown in Fig. 8.

Because the backgrounds are such a small contribution, there is no power in the fit to determine their level, and their normalizations cannot be floated. Therefore sideband constraints developed for previous analyses are used. These include *ex situ* assays of activity levels, fits in sidebands, and constrained modeling [23]. A summary of the backgrounds and constraints used in the fit are summarized in Table IV.

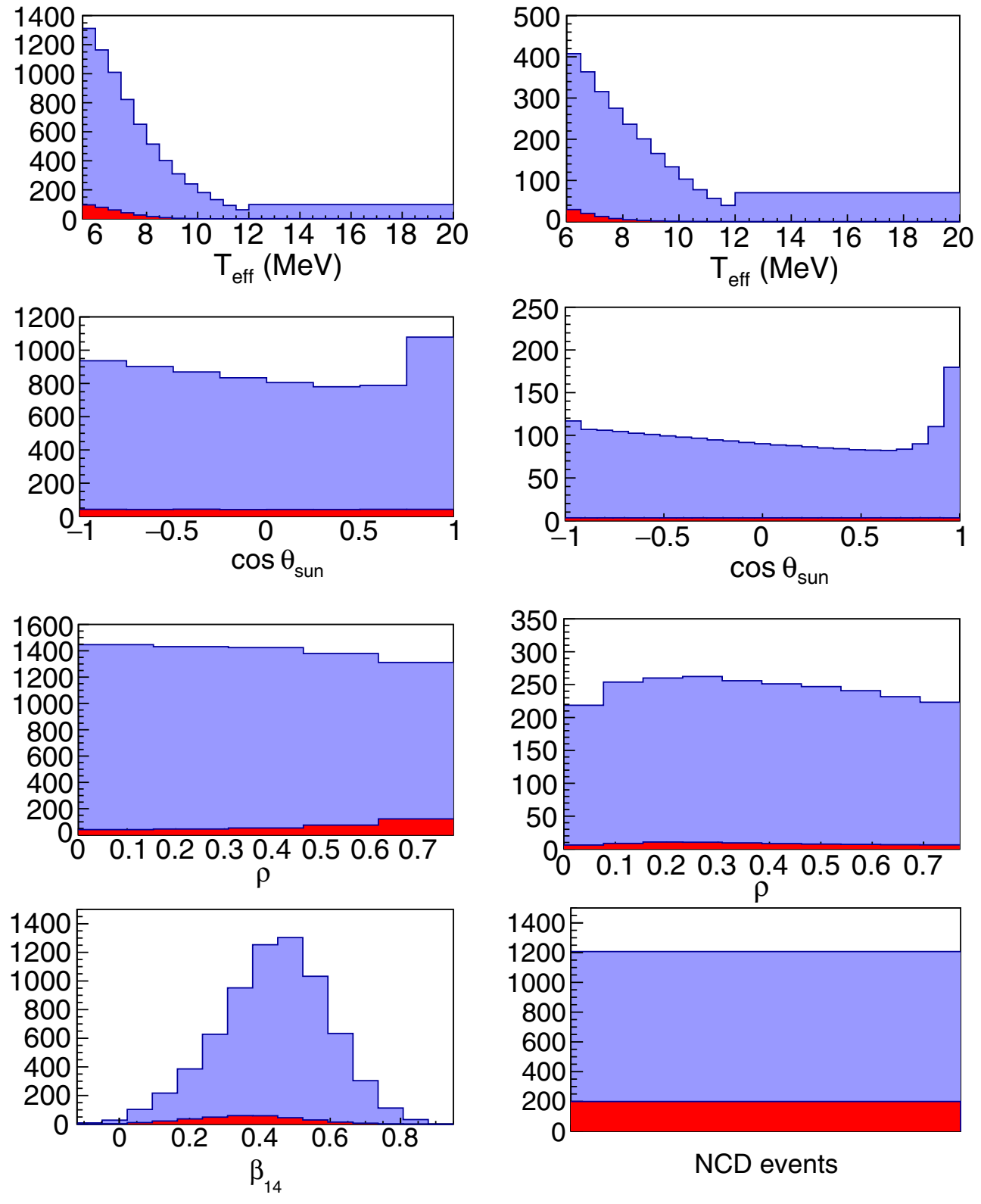

FIG. 8. Estimated contributions from signal (blue) and background (red) events within the analysis region. The distributions are taken from Monte Carlo. The left column is a sum of Phases I and II, with panels (from top) showing energy, solar angle, radius, and $\beta_{14}$. The right column is Phase III with panels (from top) showing energy, solar angle, radius, and NCD data.

### E. Bias and pull testing

We generated fake data samples from the Monte Carlo, weighted according to the neutrino oscillation model being used with an option to include Lorentz symmetry violating effects if desired. Two ensembles of 100 such fake data samples were generated, one group with no Lorentz violations present, the other with Lorentz violation at the level of $a^{(3)}_{\rm SNO} = 3\ \mathrm{GeV}^{-1}$.

The fit was run on these samples and the distribution of results examined. We found the results, summarized in Table V, to be completely consistent with an unbiased result, and the pulls to have appropriate widths.

### F. Systematics

Before unblinding the data, we evaluated which if any of the systematic uncertainties were likely to contribute significantly to the final limit.

TABLE IV. Background levels and constraints. The overall normalizations are taken from Appendix D of Ref. [24], scaled by the MC ratio of the acceptances of the different energy thresholds. The uncertainty comes from the indicated source. Phase IIIb refers to the NCD data, while Phase III refers to Cherenkov light data. K2 and K5 refer to specific NCD strings that were observed to be hotter than the rest of the array. PD stands for photodissociation. The uncertainty columns show the estimated systematic uncertainty in the Lorentz violation parameter (as a fraction of the estimated statistical uncertainty) contributed by the uncertainty in the normalization of each source of background.

| Background | Phase | Constraint (events) | Uncertainty (E) | Uncertainty ($E^2$) | Source |
|---|---|---|---|---|---|
| AV neutrons | I | $1.63 \pm 0.48$ | 0.001 | 0.003 | [3] |
| Tl D2O | I | $1.67 \pm 0.76$ | 0.001 | 0.003 | [21] |
| Bi D2O | I | $0.91 \pm 0.30$ | 0.001 | 0.003 | [21] |
| Atmospherics | I | $5.51 \pm 1.03$ | 0.001 | 0.003 | [21] |
| Tl h2o | I | $0.46 \pm 0.15$ | 0.001 | 0.000 | [21] |
| Bi h2o | I | $0.00 \pm 0.01$ | 0.001 | 0.000 | [21] |
| Tl AV | I | $0.50 \pm 0.50$ | 0.001 | 0.000 | |
| Bi AV | I | $0.08 \pm 0.08$ | 0.001 | 0.000 | |
| AV neutrons | II | $27.21 \pm 9.39$ | 0.003 | 0.003 | [25] |
| Tl D2O | II | $22.89 \pm 13.20$ | 0.003 | 0.003 | [21] |
| Bi D2O | II | $16.22 \pm 9.57$ | 0.003 | 0.003 | [25] |
| Atmospherics | II | $7.62 \pm 1.46$ | 0.001 | 0.003 | [21] |
| Tl h2o | II | $3.70 \pm 1.15$ | 0.001 | 0.003 | [21] |
| Bi h2o | II | $1.54 \pm 0.37$ | 0.001 | 0.000 | [21] |
| Tl AV | II | $6.62 \pm 6.62$ | 0.002 | 0.003 | |
| Bi AV | II | $1.85 \pm 1.85$ | 0.002 | 0.003 | |
| Na24 | II | $2.58 \pm 0.63$ | 0.001 | 0.003 | [21] |
| Atmospherics | III | $6.25 \pm 1.24$ | 0.001 | 0.000 | [21] |
| Ext n | III | $4.47 \pm 2.25$ | 0.001 | 0.000 | [21] |
| K2 | III | $2.97 \pm 0.48$ | 0.001 | 0.000 | [21] |
| K5 | III | $3.61 \pm 0.66$ | 0.001 | 0.000 | [21] |
| D2O PD | III | $2.10 \pm 0.32$ | 0.001 | 0.000 | [21] |
| NCD PD | III | $1.79 \pm 0.61$ | 0.001 | 0.000 | [21] |
| Atmospherics | IIIb | $13.60 \pm 2.70$ | 0.001 | 0.000 | [21] |
| Ext n | IIIb | $40.90 \pm 20.60$ | 0.001 | 0.003 | [21] |
| K2 | IIIb | $32.80 \pm 5.30$ | 0.001 | 0.000 | [21] |
| K5 | IIIb | $45.50 \pm 8.40$ | 0.001 | 0.000 | [21] |
| D2O PD | IIIb | $31.00 \pm 4.70$ | 0.001 | 0.000 | [21] |
| NCD PD | IIIb | $35.60 \pm 12.17$ | 0.000 | 0.000 | [21] |

TABLE V. Results of bias and pull testing. LV bias reported in units of $\mathrm{GeV}^{-1}$; flux bias reported in units of $10^6\ \mathrm{cm}^{-2}\ \mathrm{s}^{-1}$.

| Parameter | Mean | RMS |
|---|---|---|
| LV bias | $0.005 \pm 0.04$ | $0.35 \pm 0.03$ |
| LV pull | $0.0 \pm 0.1$ | $1.2 \pm 0.1$ |
| $\sin\theta_{12}$ bias | $0.0004 \pm 0.0010$ | $0.0095 \pm 0.0007$ |
| $\sin\theta_{12}$ pull | $0.028 \pm 0.1$ | $0.98 \pm 0.07$ |
| Flux bias | $-0.0021 \pm 0.012$ | $0.12 \pm 0.01$ |
| Flux pull | $-0.02 \pm 0.1$ | $1.01 \pm 0.08$ |

TABLE VI. Systematic errors arising from uncertainty on PDF shapes assuming a low energy threshold of 7.0 MeV. Errors are expressed as a fraction of the statistical $1 - \sigma$ uncertainty, which is 0.4 $\mathrm{GeV}^{-1}$ for the linear terms and 40 $\mathrm{GeV}^{-2}$ for the quadratic terms.

| Effect | Constraint | Error (E) | Error ($\mathrm{E}^2$) |
|---|---|---|---|
| E scale (3 phases) | 0.0041 | 0.004 | 0.006 |
| E scale (Phase I) | 0.0039 | 0.007 | 0.003 |
| E scale (Phase II) | 0.0034 | 0.008 | 0.003 |
| E scale (Phase III) | 0.0081 | 0.005 | 0.011 |
| E nonlin (3 phases) | 0.0069 | 0.002 | 0.006 |
| E resol (Phase I) | 0.041 | 0.006 | 0.003 |
| E resol (Phase II e) | 0.041 | 0.006 | 0.003 |
| E resol (Phase II n) | 0.018 | 0.006 | 0.003 |
| $\beta_{14}$ scale (Phase I) | 0.0042 | 0.006 | 0.003 |
| $\beta_{14}$ scale (Phase II e) | 0.0024 | 0.006 | 0.003 |
| $\beta_{14}$ scale (Phase II n) | 0.0038 | 0.006 | 0.003 |
| Dir scale (Phase III) | 0.12 | 0.006 | 0.003 |
| n Eff (Phase III) | 0.028 | 0.006 | 0.003 |
| n Eff (Phase IIIb) | 0.024 | 0.006 | 0.003 |
| $\Delta m^2_{12}$ | 0.024 | 0.007 | 0.003 |
| $\Delta m^2_{23}$ | 0.036 | 0.007 | 0.003 |
| Bkg time var | 0.5 | 0.255 | 0.255 |
| Neutrino hierarchy | | 0.006 | 0.003 |
| Earth matter pot | 0.1 | 0.006 | 0.003 |
| Seasonal day-night | | 0.030 | 0.086 |

We allowed the normalizations of each of the backgrounds listed in Table IV to vary. The estimated systematic uncertainty associated with each is shown. As can be seen, these all have a negligible impact. This can be understood from the fact that adding a few events spread evenly over the course of the experiment does very little to mimic a time-varying signal.

In addition, we considered 20 effects that control the shapes of the various PDFs used in the fit. These effects, summarized in Table VI, also have quite small effects in general.

Among the effects considered, the most concerning is the possibility that the rate of background events varies in time. Since this has the potential to mimic a signal, it is important to provide a constraint on such variations. Our strategy for handling this issue is discussed next.

Some backgrounds are considered implausible to have varied over time, such as the rates of radioactive backgrounds coming from the AV, which had no mechanism for changing. The background sources for which changes are considered plausible can be classified conveniently into externals (from the light water), internals (from the heavy water), and cosmics.

Time variations in atmospheric neutrino backgrounds cannot be studied *in situ* given the very low statistics available; however, the generation of atmospherics is well-understood, and at the relevant energies, the annual variations are very modest [26].

For the internal and external radioactivity sources, we defined sidebands in which to investigate the degree to which these backgrounds are stable in time. To study external backgrounds, we used a sideband selected by applying the following cuts: (1) $6.19\ \mathrm{m} < r < 7.02\ \mathrm{m}$, (2) $3.5\ \mathrm{MeV} < T_{\mathrm{eff}} < 20\ \mathrm{MeV}$, (3) outward-going reconstructed momentum, (4) $-0.12 < \beta_{14} < 2$ (PMT anisotropy), and (5) $0.55 <$ ITR (fraction of prompt hits).

This selection consists of roughly 90% light water external background events, with roughly 5% each of PMT and AV backgrounds that must remain constant in time. The events were binned by day and normalized by the livetime. The result was fitted to a constant plus a time-varying term according to

$$y = A(1 + B\sin(\omega t + \phi)). \tag{18}$$

Here $A$ (overall normalization) and $B$ (fractional power in the particular mode in question) are allowed to float. $\omega$ is set for either a once- or twice-annual cycle, and $\phi$ is either 0 or $\pi/2$. The power observed in the relevant Fourier modes is summarized in Table VII. The data for Phases I and II are shown in Fig. 9.

To convert the variation levels reported in Table VII into final numbers used to estimate the systematic uncertainty of the result, they must be scaled by the fraction of events of interest (external radioactivity) in the sample, about 0.9. We therefore arrive at an estimate of 15% variations in the external radioactivity levels on the timescales of interest during Phases I and II.

For Phase III, the analysis was complicated by the fact that we do not have a reliable energy reconstruction for events that spatially reconstruct outside the fiducial volume. We therefore could not apply the same energy cut directly. As a proxy for an energy cut, we applied an $N_{\mathrm{hit}}$ cut at 24. $N_{\mathrm{hit}}$ is the number of inward-looking PMTs observing a signal during the trigger window. This threshold was determined by looking at Monte Carlo simulations of external events during Phase III. It was expected that this cut would vary over time, but an examination of internal backgrounds in Phase III showed that this was not necessary.

TABLE VII. Best fit for variations in relevant Fourier modes for the different sideband samples.

| | External backgrounds | | Internal backgrounds | |
|---|---|---|---|---|
| Mode | Phases I and II | Phase III | Phases I and II | Phase III |
| $\sin\omega t$ | $-9.2\% \pm 0.5\%$ | $34.9\% \pm 0.3\%$ | $-4.2\% \pm 1.2\%$ | $0.9\% \pm 0.9\%$ |
| $\cos\omega t$ | $-12.3\% \pm 0.5\%$ | $-49.4\% \pm 0.3\%$ | $5.6\% \pm 1.3\%$ | $-4.7\% \pm 0.9\%$ |
| $\sin 2\omega t$ | $-12.5\% \pm 0.5\%$ | $3.7\% \pm 0.3\%$ | $10.0\% \pm 1.3\%$ | $-5.8\% \pm 0.9\%$ |
| $\cos 2\omega t$ | $-4.5\% \pm 0.5\%$ | $7.5\% \pm 0.3\%$ | $-9.2\% \pm 1.3\%$ | $3.9\% \pm 0.9\%$ |

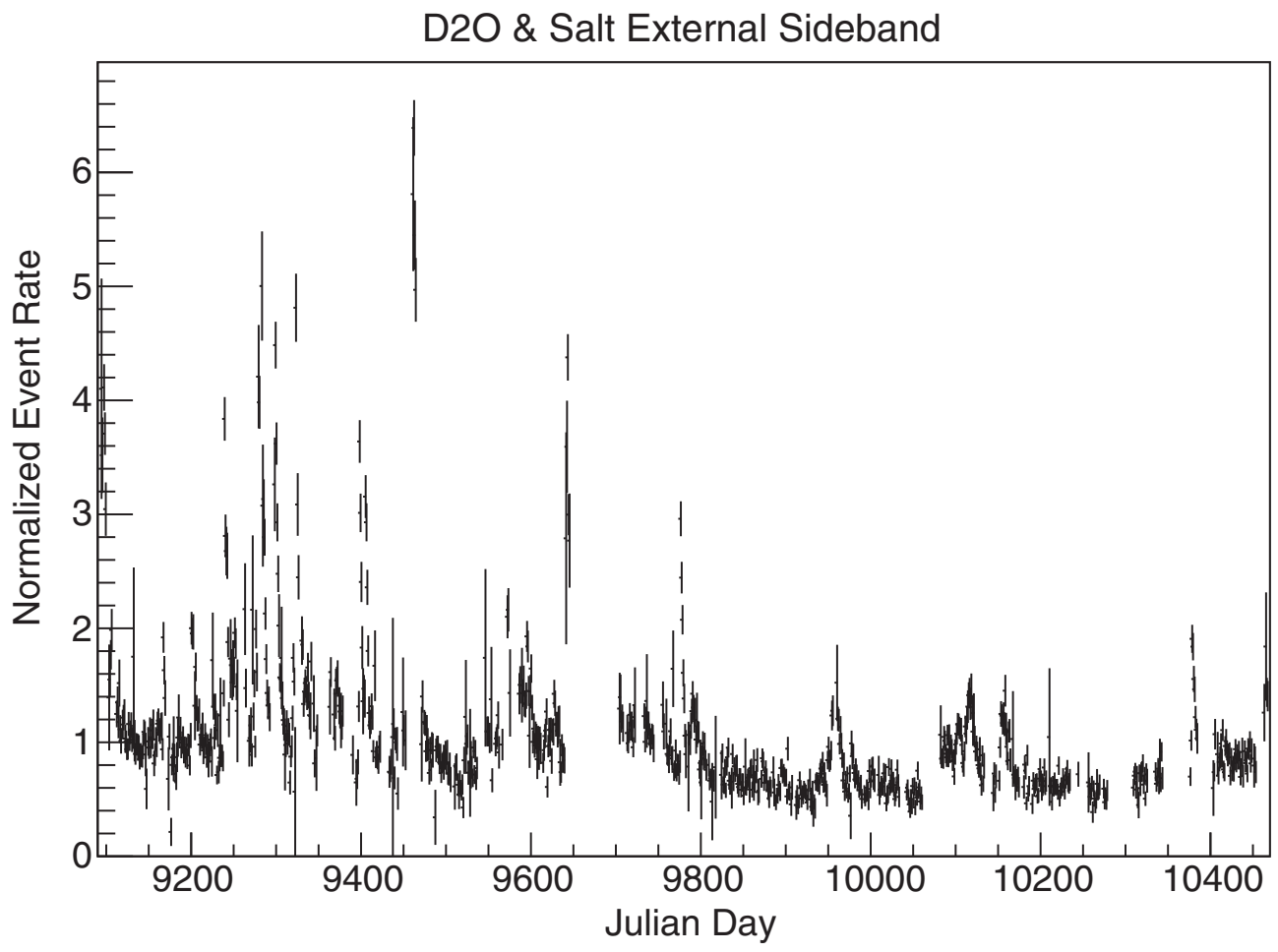

FIG. 9. Ratio of the number of events observed each day to the expected number in the light water background sideband for Phases I and II. On multiple occasions, the background level rose quickly before gradually dropping as the water was cleaned through recirculation.

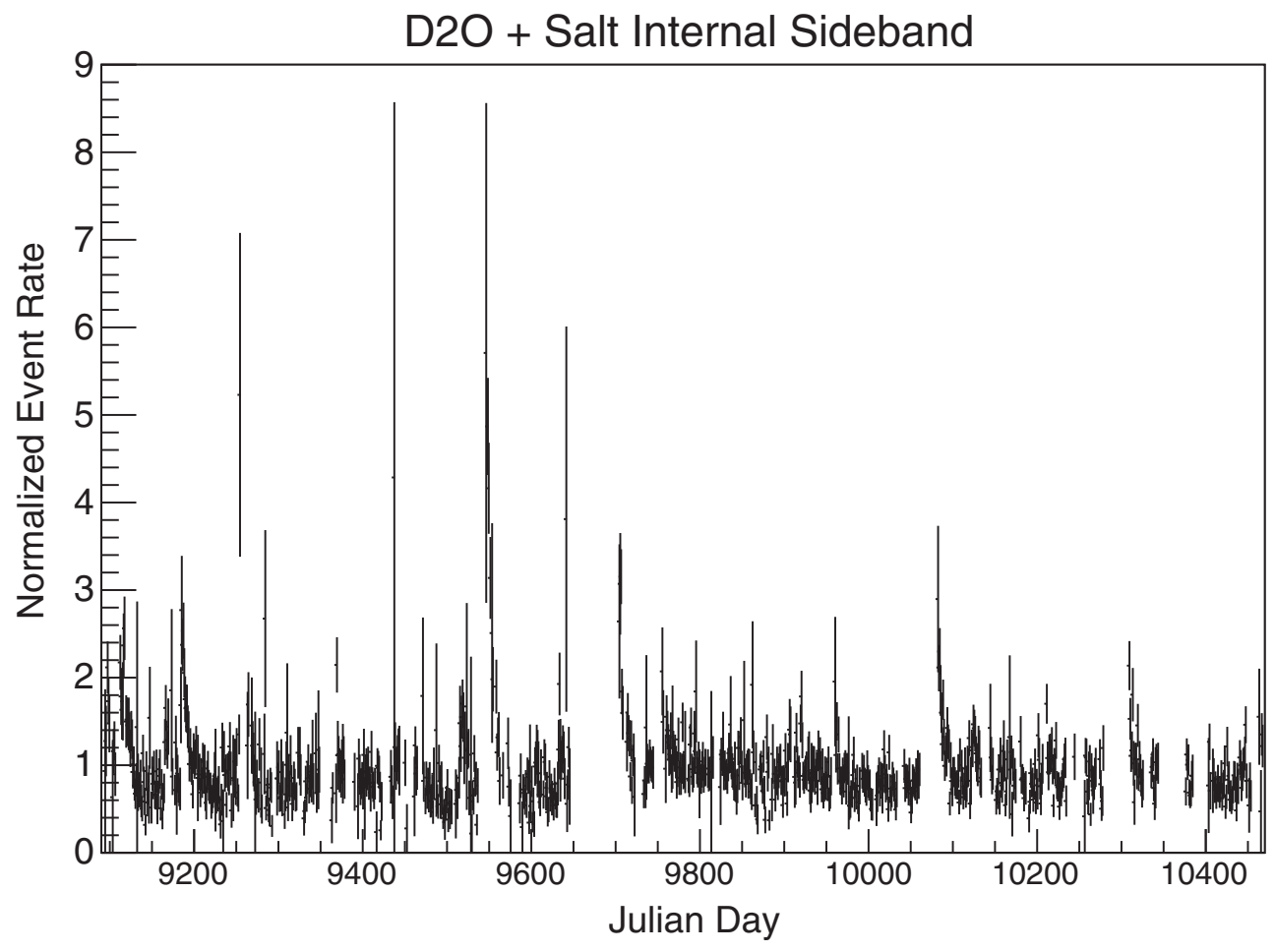

FIG. 11. Ratio of the number of events observed each day to the expected number in the internal low energy background sideband during Phases I and II. Internal backgrounds proved to be significantly more stable than external backgrounds.

Aside from this change from an energy cut to an $N_{\mathrm{hit}}$ cut, the Phase III data were handled in the same way as for the first two phases. A plot of the data is shown in Fig. 10. These data were also fit with the same kind of oscillatory model, Eq. (18). The best fit for each of the relevant modes is shown in Table VII. Because we were using $N_{\mathrm{hit}}$ as a proxy for energy, we checked whether the results depended strongly on the exact value of the $N_{\mathrm{hit}}$ cut that was used. We found that the results were very robust to such changes, changing by no more than a few percent when changing the cut by up to two hits in either direction. The large (roughly 50%) variations observed in the external backgrounds in Phase III ultimately dominated the systematic uncertainty and helped to motivate the choice of energy threshold.

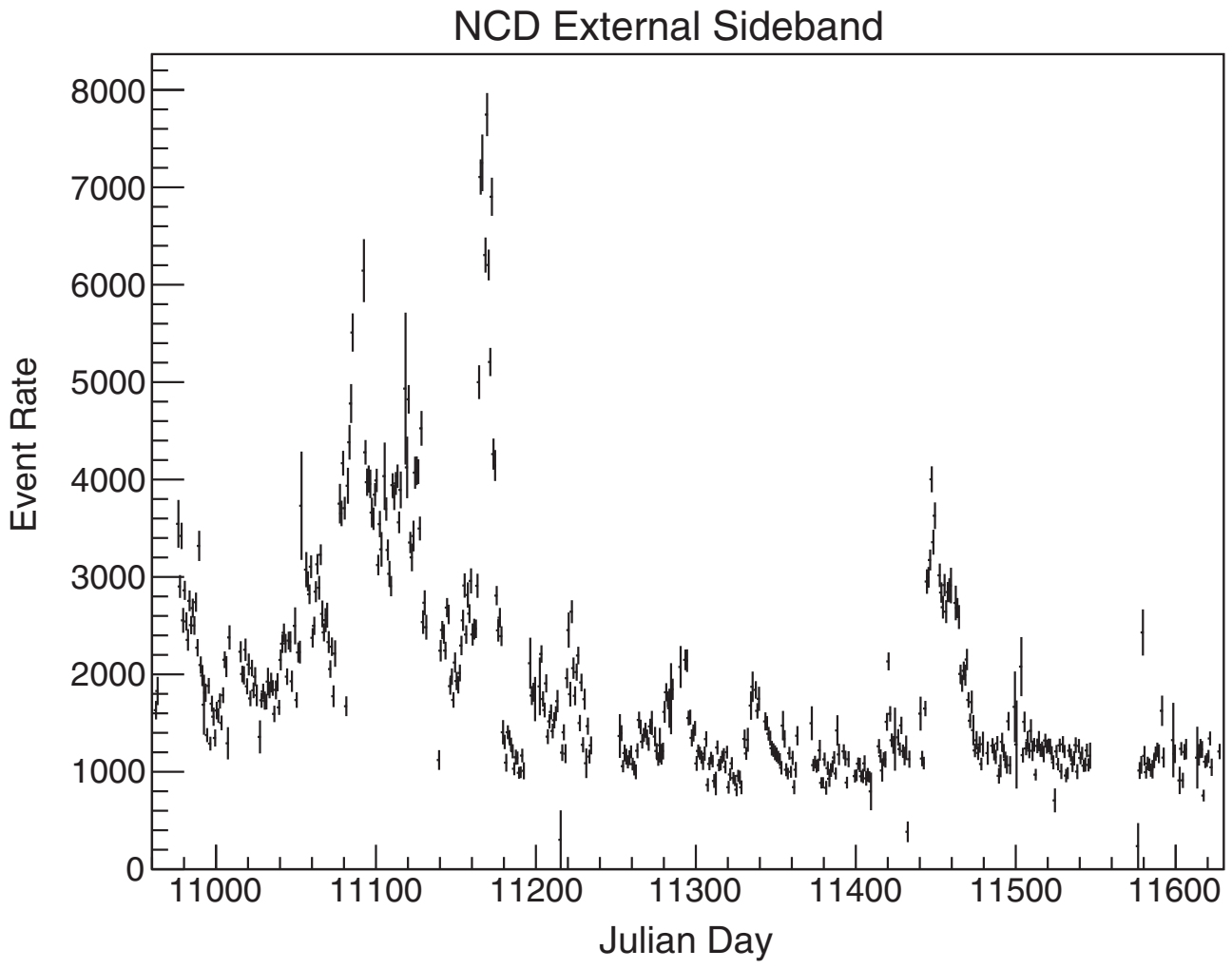

FIG. 10. Livetime-normalized events observed each day in the light water background sideband during Phase III. The large excursions observed dominate our systematic uncertainty.

For internal backgrounds, we defined an energy sideband, accepting events with reconstructed energies $3.5\ \mathrm{MeV} < T_{\mathrm{eff}} < 5.5\ \mathrm{MeV}$ and with all other cuts as in the main analysis.

Since it is not possible to isolate a pure sample of internal backgrounds (this particular selection is roughly 50% internals), the fitting procedure for this case was somewhat more complex than for the external sideband. We used a model in which the solar events were assumed to be fixed at the expected (oscillated) rate, and the backgrounds were allowed to float with the form of Eq. (18).

A plot of the data is shown in Fig. 11. The best fit for the fractional variation ($B$) in each mode, and the overall normalization of the background rate ($A$) relative to the nominal expected rate are summarized in Table VIII. Since the constraints on the backgrounds are all greater than 10% at $1\sigma$, this level of agreement is quite satisfactory.

TABLE VIII. Best fits for the various Fourier modes of the internal backgrounds during Phases I and II.

| Mode | Background variation | Background norm. |
|---|---|---|
| $\sin\omega t$ | $-4.2 \pm 1.2\%$ | $91.9 \pm 0.8\%$ |
| $\cos\omega t$ | $5.6 \pm 1.3\%$ | $91.6 \pm 0.9\%$ |
| $\sin 2\omega t$ | $10.0 \pm 1.3\%$ | $91.5 \pm 0.9\%$ |
| $\cos 2\omega t$ | $-9.2 \pm 1.3\%$ | $91.7 \pm 0.8\%$ |

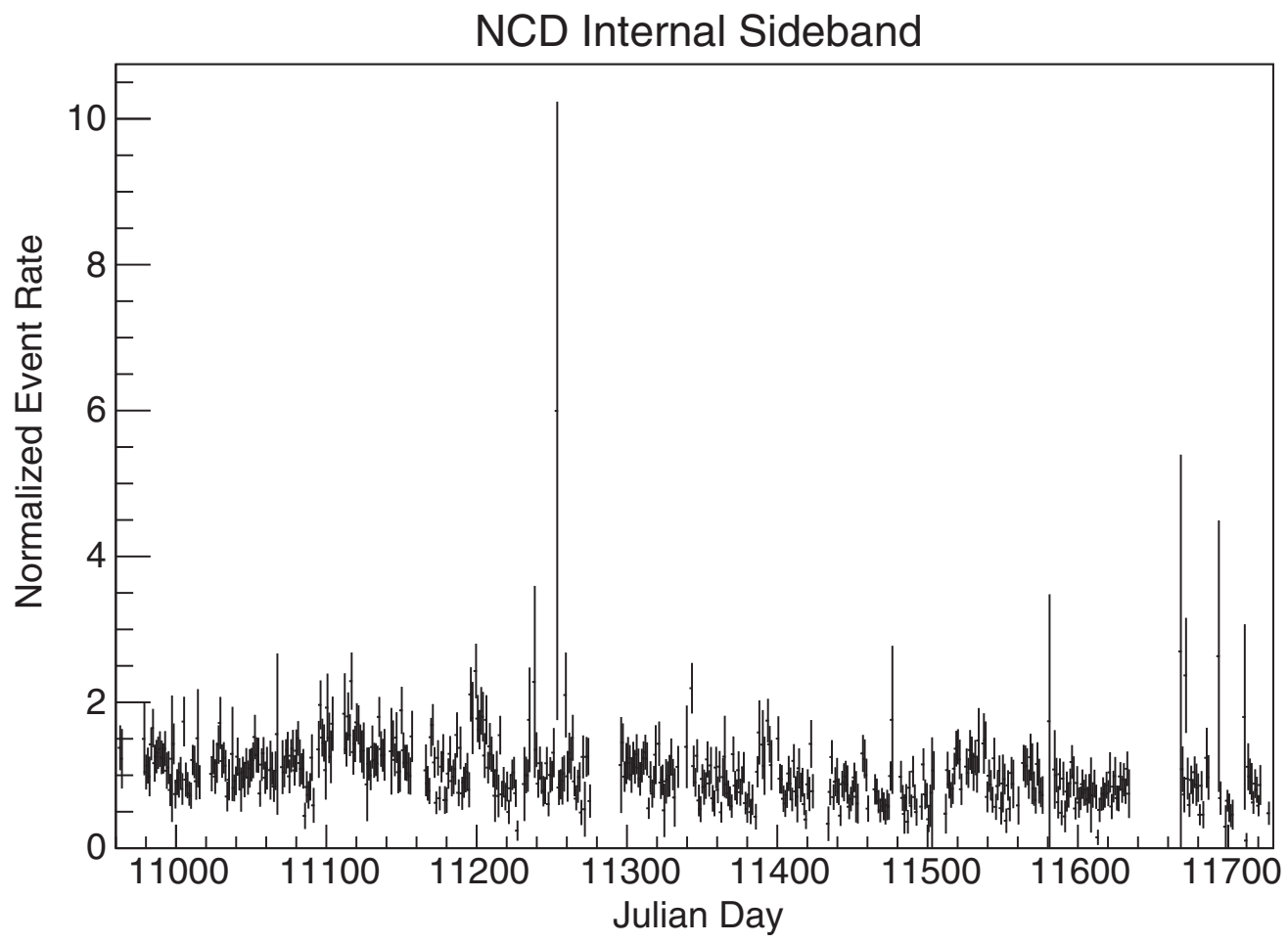

FIG. 12. Ratio of the number of events observed per day to the expected number in the internal low energy background sideband during Phase III. Internal backgrounds were significantly more stable than external backgrounds during Phase III.

We used the same procedure for the internal backgrounds for Phase III as well. A plot of the data is shown in Fig. 12 with the best-fit values of the relevant variations shown in Table VII. Note that the background normalization is not recorded for this case because Monte Carlo simulations of the low-energy backgrounds were not available for Phase III.

Since none of these systematic effects contributes an uncertainty approaching the expected statistical uncertainty of the measurement, we decided to treat the systematics through a shift-and-refit procedure. We fit the data using PDFs generated with the systematic values perturbed away from their central value by random Gaussian-distributed amounts in all dimensions simultaneously. The RMS of the distribution of fit results using these perturbed PDFs was then taken to represent the systematic uncertainty of the measurement. This technique is attractive because it automatically captures correlations between the impact of the different systematics on the final result.

## VI. RESULTS

The best-fit results for each of the eight modes are shown in Table IX. The projections of the fit are shown in Fig. 13. The reduced $\chi^2$ (11943/8765, p = 0.01) is dominated by a single event in an unlikely bin. Neglecting that bin, the reduced $\chi^2$ is 1.02 (8905/8764, p = 0.30). The data appear uniform and noiselike across all three phases, see Fig. 14. There is a hint (see Table X) of short-term variations in total event rate, particularly in the data binned at 5-day intervals, possibly due to changing background levels or detector conditions, but these effects appear to wash out on seasonal time scales.

To determine limits on the individual flavor components of the Lorentz violation effects, the limits on the different time-dependent modes shown in Table IX must be combined with information about the weight coefficients, as can be seen from Eq. (15). Since the weights depend on the mixing angles, in principle they should be recalculated for each fit mode. However, among the seven fits there were only three distinct best-fit values for the solar mixing angle. We therefore calculated the weights for each of these three cases, as shown in Table XI.

The results in Tables IX and XI cannot simply be divided to attain the limits on the flavor components because the weights share common systematic uncertainties with the Lorentz violation signal fit results (for example, the value of $\theta_{13}$). To correctly account for these correlations, we calculated the limit on the signal and the weight for each member of the ensemble and determined the RMS of their ratio.

The limits on the various individual flavor components (assuming the others are zero) are listed in Tables XII and XIII. We set limits on 38 previously unconstrained parameters and set improved limits on 16 additional parameters. For the first time, limits are now available on every leading-order Lorentz violation operator in the neutrino sector.

### A. Interpretation as energy scale

It is expected that if Lorentz symmetry violations exist in nature, they would derive from new physics at a high

TABLE IX. Lorentz violation best-fit results. The first error is statistical and the second systematic.

| Mode | LV signal | Solar flux ($10^6$ cm$^{-2}$ s$^{-1}$) | $\sin\theta_{12}$ |
|---|---|---|---|
| $E$ | $7.0^{+7.2+5.9}_{-7.5-6.7}$ GeV$^{-1}$ | $5.22 \pm 0.27^{+0.17}_{-0.22}$ | $0.497\ ^{+0.088+0.078}_{-0.098-0.078}$ |
| $E\sin\omega t$ | $0.0^{+7.2+2.1}_{-7.3-2.2} \times 10^{-1}$ GeV$^{-1}$ | $5.15 \pm 0.26^{+0.14}_{-0.17}$ | $0.577\ ^{+0.019+0.010}_{-0.018-0.009}$ |
| $E\cos\omega t$ | $0.2^{+7.3+2.2}_{-7.4-2.3} \times 10^{-1}$ GeV$^{-1}$ | $5.15 \pm 0.26^{+0.14}_{-0.17}$ | $0.577\ ^{+0.019+0.010}_{-0.018-0.009}$ |
| $E^2$ | $3.0^{+3.3+2.7}_{-3.4-3.1} \times 10^2$ GeV$^{-2}$ | $5.22 \pm 0.27^{+0.17}_{-0.22}$ | $0.537\ ^{+0.048+0.042}_{-0.049-0.037}$ |
| $E^2\sin\omega t$ | $0.7^{+6.4+1.7}_{-6.5-1.8} \times 10^1$ GeV$^{-2}$ | $5.15 \pm 0.26^{+0.14}_{-0.17}$ | $0.577\ ^{+0.019+0.011}_{-0.018-0.008}$ |
| $E^2\cos\omega t$ | $-0.2^{+6.5+1.9}_{-6.6-1.9} \times 10^1$ GeV$^{-2}$ | $5.15 \pm 0.26^{+0.14}_{-0.17}$ | $0.577\ ^{+0.019+0.010}_{-0.018-0.009}$ |
| $E^2\sin 2\omega t$ | $5.8^{+6.5+1.6}_{-6.4-1.8} \times 10^1$ GeV$^{-2}$ | $5.15 \pm 0.26^{+0.14}_{-0.17}$ | $0.577\ ^{+0.019+0.010}_{-0.018-0.009}$ |
| $E^2\cos 2\omega t$ | $-4.4^{+6.5+1.7}_{-6.6-1.8} \times 10^1$ GeV$^{-2}$ | $5.15 \pm 0.26^{+0.14}_{-0.17}$ | $0.577\ ^{+0.019+0.010}_{-0.018-0.009}$ |

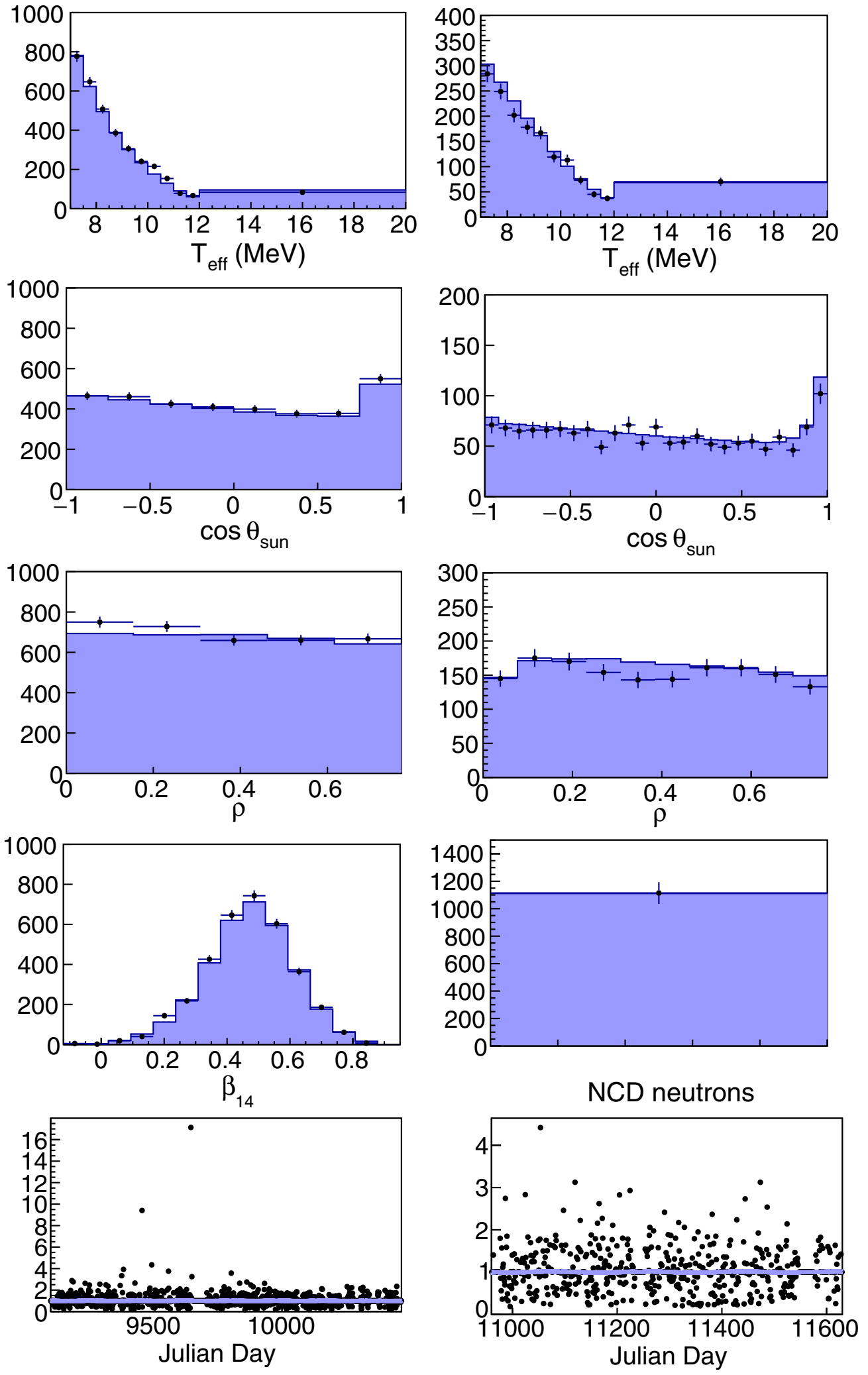

FIG. 13. Fit for $E^2 \sin 2\omega t$ term projected along the various axes. Data shown as points with poisson errors; filled histogram shows best fit. The left column shows data for Phases I and II; the right column is for Phase III. The bottom row shows the data in black points with the fit result in blue. Since the daily residuals are difficult to see, they are shown rebinned on longer timescales in Fig. 14.

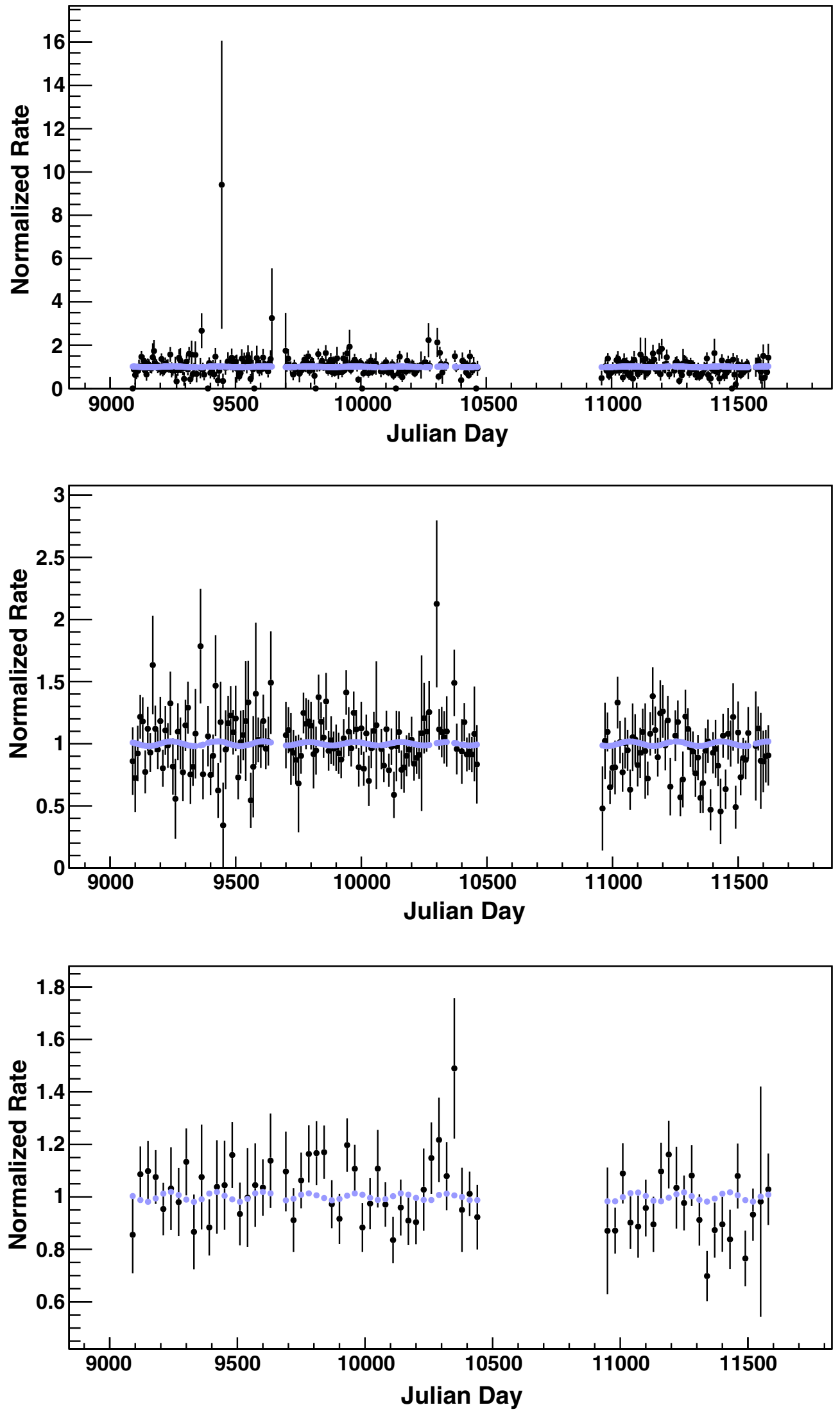

FIG. 14. Time residuals of the fit of the full data set, binned over 5 (top), 10 (middle), and 30 (bottom) day periods. The blue points indicate the shape of the signal for the best fit.

energy scale. In the simplest cases, effects at low energies would be suppressed by a factor of

$$g \frac{m_{\mathrm{EW}}}{m_{\mathrm{NP}}} \tag{19}$$

relative to electroweak physics [5], where $g$ is a coupling constant, $m_{\mathrm{EW}} \approx 100$ GeV is the electroweak mass scale, and $m_{\mathrm{NP}}$ is the mass scale of the new physics. This provides a kind of benchmark for evaluating the reach of the limits established here.

Assuming "natural" models should have couplings no smaller than 0.01, the limits we set here rule out models of this kind up to mass scales of the order of $10^{17}$ GeV. Of course, more complex kinds of models can evade such limits [31].

## B. Comparison to previous SNO analyses

As a cross check of this analysis, we ran a fit in which Lorentz violations were constrained to be zero. In this configuration, the fit results for the solar flux and mixing

TABLE X. $\chi^2$/ndf for the time residuals binned on different time scales. Although there is a hint of short-period changes to the total event rate, these effects average away on seasonal time scales. $p$ shows the p-value for rejecting the hypothesis of the data being constant in time.

| Binning period | $\chi^2$/ndf | p |
|---|---|---|
| 5 days | 390/340 | 0.07 |
| 10 days | 203/188 | 0.23 |
| 30 days | 68/66 | 0.44 |

TABLE XI. Estimates for the weight coefficients in units of $10^{-2}$ $\mathrm{meV}^{-2}$ ($10^{22}$ $\mathrm{GeV}^{-2}$). The first error comes from the uncertainty of the best-fit result; the second error is systematic.

| Flavor | Time fits | E | $E^2$ |
|---|---|---|---|
| ee | $0.289 \pm 0.013 \pm 0.044$ | $0.230 \pm 0.082 \pm 0.038$ | $0.261 \pm 0.043 \pm 0.042$ |
| $e\mu$ | $-0.263 \pm 0.047 \pm 0.026$ | $-0.427 \pm 0.250 \pm 0.045$ | $-0.347 \pm 0.138 \pm 0.036$ |
| $e\tau$ | $-0.393 \pm 0.003 \pm 0.098$ | $-0.392 \pm 0.005 \pm 0.040$ | $-0.394 \pm 0.005 \pm 0.039$ |
| $\mu\mu$ | $-0.232 \pm 0.009 \pm 0.038$ | $-0.189 \pm 0.059 \pm 0.031$ | $-0.212 \pm 0.031 \pm 0.035$ |
| $\mu\tau$ | $-0.257 \pm 0.034 \pm 0.045$ | $-0.121 \pm 0.196 \pm 0.039$ | $-0.189 \pm 0.105 \pm 0.042$ |
| $\tau\tau$ | $-0.057 \pm 0.004 \pm 0.007$ | $-0.041 \pm 0.023 \pm 0.007$ | $-0.042 \pm 0.019 \pm 0.014$ |

angle agreed within $1\sigma$ with the previous results of the full SNO data set published in [21]. An analysis accounting for the nonindependence of the two samples showed no significant evidence of disagreement between the two analyses. The full analysis of all three phases of SNO data, reported in [21], used a significantly larger

TABLE XII. Comparison to existing limits for $a$ coefficients. All $a$'s here refer to $a_{\mathrm{eff}}^{(3)}$. All entries without a previous limit noted were not previously constrained. Limits marked * are technically set on $a_L$, but can be interpreted as limits on $a_{\mathrm{eff}}$ for reasons already stated. Results marked † are stated at $3\sigma$ CL, all others at 95% CL. Information collected from [10].

| Coefficient | This work | Previous limit | Ref. |
|---|---|---|---|
| $\lvert a_{00}^{ee}\rvert$ | $8.8\times10^{-20}$ GeV | | |
| $\lvert a_{00}^{e\mu}\rvert$ | $9.8\times10^{-20}$ GeV | $9.2\times10^{-20}$ GeV | [11] |
| $\lvert a_{00}^{e\tau}\rvert$ | $6.5\times10^{-20}$ GeV | $2.8\times10^{-19}$ GeV [*] | [13] |
| $\lvert a_{00}^{\mu\mu}\rvert$ | $8.2\times10^{-20}$ GeV | $2.2\times10^{-23}$ GeV [*] | [27] |
| $\lvert a_{00}^{\mu\tau}\rvert$ | $7.5\times10^{-20}$ GeV | $5.0\times10^{-24}$ GeV [*] | [27] |
| $\lvert a_{00}^{\tau\tau}\rvert$ | $2.7\times10^{-19}$ GeV | $2.2\times10^{-23}$ GeV [*] | [27] |
| $\lvert a_{10}^{ee}\rvert$ | $4.3\times10^{-21}$ GeV | | |
| $\lvert a_{10}^{e\mu}\rvert$ | $4.2\times10^{-21}$ GeV | $7.1\times10^{-20}$ GeV | [11] |
| $\lvert a_{10}^{e\tau}\rvert$ | $2.8\times10^{-21}$ GeV | $5.5\times10^{-19}$ GeV [*] | [13] |
| $\lvert a_{10}^{\mu\mu}\rvert$ | $5.4\times10^{-21}$ GeV | | |
| $\lvert a_{10}^{\mu\tau}\rvert$ | $5.1\times10^{-21}$ GeV | $1.9\times10^{-18}$ GeV [*] | [28] |
| $\lvert a_{10}^{\tau\tau}\rvert$ | $2.0\times10^{-20}$ GeV | | |
| $\lvert\mathfrak{Re}(a_{11}^{ee})\rvert$ | $2.3\times10^{-21}$ GeV | | |
| $\lvert\mathfrak{Re}(a_{11}^{e\mu})\rvert$ | $2.2\times10^{-21}$ GeV | $8.1\times10^{-20}$ GeV | [11] |
| $\lvert\mathfrak{Re}(a_{11}^{e\tau})\rvert$ | $1.5\times10^{-21}$ GeV | $1.3\times10^{-19}$ GeV [*] | [13] |
| $\lvert\mathfrak{Re}(a_{11}^{\mu\mu})\rvert$ | $2.9\times10^{-21}$ GeV | $6.9\times10^{-20}$ GeV [*] | [29] |
| $\lvert\mathfrak{Re}(a_{11}^{\mu\tau})\rvert$ | $2.8\times10^{-21}$ GeV | $8.8\times10^{-23}$ GeV [*†] | [14] |
| $\lvert\mathfrak{Re}(a_{11}^{\tau\tau})\rvert$ | $1.1\times10^{-20}$ GeV | | |
| $\lvert\mathfrak{Im}(a_{11}^{ee})\rvert$ | $2.5\times10^{-21}$ GeV | | |
| $\lvert\mathfrak{Im}(a_{11}^{e\mu})\rvert$ | $2.5\times10^{-21}$ GeV | $8.5\times10^{-20}$ GeV | [11] |
| $\lvert\mathfrak{Im}(a_{11}^{e\tau})\rvert$ | $1.7\times10^{-21}$ GeV | $1.3\times10^{-19}$ GeV [*] | [13] |
| $\lvert\mathfrak{Im}(a_{11}^{\mu\mu})\rvert$ | $3.2\times10^{-21}$ GeV | $6.9\times10^{-20}$ GeV [*] | [29] |
| $\lvert\mathfrak{Im}(a_{11}^{\mu\tau})\rvert$ | $3.1\times10^{-21}$ GeV | $8.8\times10^{-23}$ GeV [*†] | [14] |
| $\lvert\mathfrak{Im}(a_{11}^{\tau\tau})\rvert$ | $1.2\times10^{-20}$ GeV | | |

TABLE XIII. Comparison to existing limits for $c$ coefficients. All $c$'s refer to $c_{\mathrm{eff}}^{(4)}$. Entries without a previous limit noted were not previously constrained. Limits marked * are technically set on $c_L$, but these can be applied to $c_{\mathrm{eff}}$ for reasons already discussed. Results marked † are stated at $3\sigma$ CL, ‡ at 99% CL, all others at 95% CL. Data collected from [10].

| Coefficient | This work | Previous limit | Reference |
|---|---|---|---|
| $\lvert c_{00}^{ee}\rvert$ | $2.3\times10^{-18}$ | | |
| $\lvert c_{00}^{e\mu}\rvert$ | $2.5\times10^{-18}$ | $1.5\times10^{-19}$ | [11] |
| $\lvert c_{00}^{e\tau}\rvert$ | $1.6\times10^{-18}$ | $1.4\times10^{-16}$ [*] | [13] |
| $\lvert c_{00}^{\mu\mu}\rvert$ | $2.9\times10^{-18}$ | | |
| $\lvert c_{00}^{\mu\tau}\rvert$ | $2.7\times10^{-18}$ | $1.4\times10^{-27}$ [‡] | [30] |
| $\lvert c_{00}^{\tau\tau}\rvert$ | $1.1\times10^{-17}$ | | |
| $\lvert c_{10}^{ee}\rvert$ | $3.9\times10^{-19}$ | | |
| $\lvert c_{10}^{e\mu}\rvert$ | $3.7\times10^{-19}$ | $1.2\times10^{-19}$ | [11] |
| $\lvert c_{10}^{e\tau}\rvert$ | $2.5\times10^{-19}$ | | |
| $\lvert c_{10}^{\mu\mu}\rvert$ | $4.8\times10^{-19}$ | | |
| $\lvert c_{10}^{\mu\tau}\rvert$ | $4.5\times10^{-19}$ | | |
| $\lvert c_{10}^{\tau\tau}\rvert$ | $1.8\times10^{-18}$ | | |
| $\lvert\mathfrak{Re}(c_{11}^{ee})\rvert$ | $2.0\times10^{-19}$ | | |
| $\lvert\mathfrak{Re}(c_{11}^{e\mu})\rvert$ | $2.0\times10^{-19}$ | $1.3\times10^{-19}$ | [11] |
| $\lvert\mathfrak{Re}(c_{11}^{e\tau})\rvert$ | $1.3\times10^{-19}$ | | |
| $\lvert\mathfrak{Re}(c_{11}^{\mu\mu})\rvert$ | $2.6\times10^{-19}$ | $1.3\times10^{-20}$ [*] | [29] |
| $\lvert\mathfrak{Re}(c_{11}^{\mu\tau})\rvert$ | $2.5\times10^{-19}$ | $7.2\times10^{-24}$ [*†] | [14] |
| $\lvert\mathfrak{Re}(c_{11}^{\tau\tau})\rvert$ | $9.8\times10^{-19}$ | | |
| $\lvert\mathfrak{Im}(c_{11}^{ee})\rvert$ | $2.2\times10^{-19}$ | | |
| $\lvert\mathfrak{Im}(c_{11}^{e\mu})\rvert$ | $2.2\times10^{-19}$ | $1.4\times10^{-19}$ | [11] |
| $\lvert\mathfrak{Im}(c_{11}^{e\tau})\rvert$ | $1.5\times10^{-19}$ | | |
| $\lvert\mathfrak{Im}(c_{11}^{\mu\mu})\rvert$ | $2.8\times10^{-19}$ | $1.3\times10^{-20}$ [*] | [29] |
| $\lvert\mathfrak{Im}(c_{11}^{\mu\tau})\rvert$ | $2.7\times10^{-19}$ | $1.3\times10^{-22}$ [*†] | [14] |
| $\lvert\mathfrak{Im}(c_{11}^{\tau\tau})\rvert$ | $1.1\times10^{-18}$ | | |
| $\lvert c_{20}^{ee}\rvert$ | $1.1\times10^{-18}$ | | |
| $\lvert c_{20}^{e\mu}\rvert$ | $1.1\times10^{-18}$ | $2.0\times10^{-19}$ | [11] |
| $\lvert c_{20}^{e\tau}\rvert$ | $7.4\times10^{-19}$ | $7.8\times10^{-16}$ [*] | [13] |
| $\lvert c_{20}^{\mu\mu}\rvert$ | $1.4\times10^{-18}$ | | |
| $\lvert c_{20}^{\mu\tau}\rvert$ | $1.4\times10^{-18}$ | | |

*(Table continued)*

TABLE XIII. *(Continued)*

| Coefficient | This work | Previous limit | Reference |
|---|---|---|---|
| $\lvert c_{20}^{\tau\tau}\rvert$ | $5.4 \times 10^{-18}$ | | |
| $\lvert \mathfrak{Re}(c_{21}^{ee})\rvert$ | $6.1 \times 10^{-19}$ | | |
| $\lvert \mathfrak{Re}(c_{21}^{e\mu})\rvert$ | $6.1 \times 10^{-19}$ | $8.0 \times 10^{-20}$ | [11] |
| $\lvert \mathfrak{Re}(c_{21}^{e\tau})\rvert$ | $4.0 \times 10^{-19}$ | | |
| $\lvert \mathfrak{Re}(c_{21}^{\mu\mu})\rvert$ | $7.7 \times 10^{-19}$ | $2.0 \times 10^{-20}$ [*] | [29] |
| $\lvert \mathfrak{Re}(c_{21}^{\mu\tau})\rvert$ | $7.2 \times 10^{-19}$ | $4.5 \times 10^{-24}$ [*†] | [14] |
| $\lvert \mathfrak{Re}(c_{21}^{\tau\tau})\rvert$ | $2.9 \times 10^{-18}$ | | |
| $\lvert \mathfrak{Im}(c_{21}^{ee})\rvert$ | $6.4 \times 10^{-19}$ | | |
| $\lvert \mathfrak{Im}(c_{21}^{e\mu})\rvert$ | $6.3 \times 10^{-19}$ | $8.5 \times 10^{-20}$ | [11] |
| $\lvert \mathfrak{Im}(c_{21}^{e\tau})\rvert$ | $4.2 \times 10^{-19}$ | | |
| $\lvert \mathfrak{Im}(c_{21}^{\mu\mu})\rvert$ | $8.1 \times 10^{-19}$ | $2.0 \times 10^{-20}$ [*] | [29] |
| $\lvert \mathfrak{Im}(c_{21}^{\mu\tau})\rvert$ | $7.7 \times 10^{-19}$ | $7.1 \times 10^{-22}$ [*†] | [14] |
| $\lvert \mathfrak{Im}(c_{21}^{\tau\tau})\rvert$ | $3.1 \times 10^{-18}$ | | |
| $\lvert \mathfrak{Re}(c_{22}^{ee})\rvert$ | $2.5 \times 10^{-19}$ | | |
| $\lvert \mathfrak{Re}(c_{22}^{e\mu})\rvert$ | $2.4 \times 10^{-19}$ | $1.7 \times 10^{-17}$ | [11] |
| $\lvert \mathfrak{Re}(c_{22}^{e\tau})\rvert$ | $1.6 \times 10^{-19}$ | $1.2 \times 10^{-17}$ [*] | [13] |
| $\lvert \mathfrak{Re}(c_{22}^{\mu\mu})\rvert$ | $3.1 \times 10^{-19}$ | | |
| $\lvert \mathfrak{Re}(c_{22}^{\mu\tau})\rvert$ | $3.0 \times 10^{-19}$ | $7.8 \times 10^{-24}$ [*†] | [14] |
| $\lvert \mathfrak{Re}(c_{22}^{\tau\tau})\rvert$ | $1.2 \times 10^{-18}$ | | |
| $\lvert \mathfrak{Im}(c_{22}^{ee})\rvert$ | $2.6 \times 10^{-19}$ | | |
| $\lvert \mathfrak{Im}(c_{22}^{e\mu})\rvert$ | $2.6 \times 10^{-19}$ | $1.7 \times 10^{-17}$ | [11] |
| $\lvert \mathfrak{Im}(c_{22}^{e\tau})\rvert$ | $1.7 \times 10^{-19}$ | $1.2 \times 10^{-17}$ [*] | [13] |
| $\lvert \mathfrak{Im}(c_{22}^{\mu\mu})\rvert$ | $3.3 \times 10^{-19}$ | | |
| $\lvert \mathfrak{Im}(c_{22}^{\mu\tau})\rvert$ | $3.1 \times 10^{-19}$ | $2.2 \times 10^{-21}$ [*†] | [14] |
| $\lvert \mathfrak{Im}(c_{22}^{\tau\tau})\rvert$ | $1.2 \times 10^{-18}$ | | |

data set and a more detailed background model and should still be understood as the definitive analysis of the SNO data set.

## VII. CONCLUSION

No evidence of Lorentz symmetry violations was found in an analysis of the data from all phases of SNO. Limits were established on all minimal, Dirac-type Lorentz violating operators in the neutrino sector; of these, 38 were previously unconstrained by experiment, and improved limits were set on 16 additional parmeters. The extensive coverage of the analysis is a consequence of the use of solar neutrinos, whose flavor changes in the Sun, due to the matter effect, make them sensitive to effects in all flavor components. Since the limits are roughly at the level expected from new physics at the Planck scale, they provide strong constraints on the possible kinds of beyond standard model physics that can be predicted by future theories.

## ACKNOWLEDGMENTS

This research was supported by Canada: Natural Sciences and Engineering Research Council, Industry Canada, National Research Council, Northern Ontario Heritage Fund, Atomic Energy of Canada, Ltd., Ontario Power Generation, High Performance Computing Virtual Laboratory, Canada Foundation for Innovation, Canada Research Chairs program; US: Department of Energy Office of Nuclear Physics, National Energy Research Scientific Computing Center, Alfred P. Sloan Foundation, National Science Foundation, the Queen's Breakthrough Fund, Department of Energy National Nuclear Security Administration through the Nuclear Science and Security Consortium; United Kingdom: Science and Technology Facilities Council (formerly Particle Physics and Astronomy Research Council); Portugal: Fundação para a Ciência e a Tecnologia. We thank the SNO technical staff for their strong contributions. We thank INCO (now Vale, Ltd.) for hosting this project in their Creighton mine.